\documentclass[12pt]{article}
\usepackage{amssymb,amsmath,latexsym,graphicx,tikz,setspace,wasysym,float,subfig}
\usepackage{hyperref}

\newcommand{\half}{{\textstyle{1\over2}}}

\newcommand{\beq}{\begin{equation}}
\newcommand{\be}{\begin{equation}}
\newcommand{\ee}{\end{equation}}
\newcommand{\bea}{\begin{eqnarray}}
\newcommand{\eea}{\end{eqnarray}}

\newcommand{\pa}{\partial}
\newcommand{\pab}{\bar{\partial}}

\newcommand{\nn}{\nonumber}

\newcommand{\ti}[1]{\tilde{#1}}

\newcommand{\zb}{\bar{z}}

\numberwithin{equation}{section}
\onehalfspace
\begin{document}

\begin{titlepage}
\hbox to \hsize{\hspace*{0 cm}\hbox{\tt }\hss
    \hbox{\small{\tt }}}

\vspace{1 cm}
\centerline{\Large Operator mixing for string states in the D1-D5 CFT near the orbifold point}

\vspace{1 cm}
 \centerline{\large Benjamin A. Burrington\footnote{bburrington@troy.edu}$^\dagger\!\!$\,, Amanda W. Peet\footnote{amanda.peet@utoronto.ca}$^\ddagger\!\!$\,, and  Ida G. Zadeh\footnote{ghazvini@physics.utoronto.ca}$^\ddagger\!\!$}

\vspace{0.3cm}
\centerline{\it ${}^\dagger$Department of Chemistry and Physics, Troy University, Troy, Alabama, USA 36082.}

\vspace{0.3cm}
\centerline{\it ${}^\ddagger$Department of Physics, University of Toronto, Toronto, Ontario, Canada M5S 1A7. }

\vspace{1 cm}

\begin{abstract}

In the context of the fuzzball programme, we investigate deforming the microscopic string description of the D1-D5 system on $T^4\times S^1$ away from the orbifold point. Using conformal perturbation theory and a generalization of Lunin-Mathur symmetric orbifold technology for computing twist-nontwist correlators developed in a companion paper \cite{companion}, we initiate a program to compute the anomalous dimensions of low-lying string states in the D1-D5 superconformal field theory.
Our method entails finding four-point functions involving a string operator ${\cal{O}}$ of interest and the deformation operator, taking coincidence limits to identify which other operators mix with ${\cal{O}}$, subtracting the identified conformal family to isolate other contributions to the four-point function, finding the mixing coefficients, and iterating. For the lowest-lying string modes, this procedure should truncate in a finite number of steps.
We check our method by showing how the operator dual to the dilaton does not participate in mixing that would change its conformal dimension, as expected.
Next we complete the first stage of the iteration procedure for a low-lying string state of the form $\partial X \partial X {\bar{\partial}} X {\bar{\partial}} X$ and find its mixing coefficient. Our most interesting qualitative result is evidence of operator mixing at first order in the deformation parameter, which means that the string state acquires an anomalous dimension. After diagonalization this will mean that  anomalous dimensions of some string states in the D1-D5 SCFT must decrease away from the orbifold point while others increase.

\end{abstract}

\end{titlepage}

\tableofcontents

\section{Introduction}\label{introduction}

Soon after the 1995 discovery by Polchinski that D-branes carry Ramond-Ramond charge, Strominger and Vafa did the first-ever computation of the Bekenstein-Hawking entropy from first principles, using statistical mechanics and perturbative string theory on a system involving  $N_1$ D1-branes and $N_5$ D5-branes. The entropy agreement was underpinned by a supersymmetric non-renormalization theorem, which does not persist in black hole systems at finite temperature or in classes of black holes with less supersymmetry.
The perturbative string description of the D1-D5 system, a conformal field theory (CFT), is valid when perturbative open and closed string corrections are small: $g_s N_{1,5}\,{\ll}\,1$ and $g_s^2 N_p\,{\ll}\,1$. Black hole spacetime calculations are reliable in the opposite limit for macroscopic quantum numbers. Therefore, to honestly account for the degrees of freedom underlying the Bekenstein-Hawking entropy of black holes, we must better understand how the two descriptions mesh together.
In general, calculating a physical quantity in the CFT does not necessarily give the same answer as the classical spacetime. That said, for certain near-extremal configurations, beautiful agreement has been demonstrated between spectra for emission from the spacetime ergoregion and from the microscopic CFT description.
A key physics point is that getting the entropy or the emission spectrum right does not necessarily mean that you get the entanglement or the quantum states right. To resolve the information paradox, morally speaking, you need to know about the wavefunction inside the black hole horizon as well as outside it.

The fuzzball program pioneered by Mathur, reviewed in e.g.~\cite{Mathur:2005zp,Bena:2007kg,Skenderis:2008qn,Balasubramanian:2008da, Mathur:2009hf, Chowdhury:2010ct, Mathur:2011uj}, is in our view our best hope for understanding in a technically detailed way where Hawking might have gone wrong nearly forty years ago in predicting that black holes destroy information. An essential part of the program is Mathur's striking result \cite{Mathur:2009hf} that only a very small fraction of the rising entanglement between traditional classical black holes and their Hawking radiation can be removed by perturbative quantum corrections. Fuzzballs finagle traditional no-hair intuition by having hair that is nonperturbative, {\em i.e.}~not explained by linearized perturbation theory about black holes. There is also an exponentially large phase space of fuzzballs with macroscopic quantum numbers. According to Mathur's conjecture, black holes arise from thermal averaging over fuzzballs: smooth gravitational duals of string theory microstates with the same conserved quantum numbers as the black hole, animals which do not possess event horizons, curvature singularities, or spherical symmetry. Reconciling the fuzzball picture with GR intuition about infall is an active area of investigation; see e.g.~\cite{Mathur:2012jk, Chowdhury:2012vd, Bena:2012zi, Avery:2012tf, Lunin:2012gz}.

So far, the majority of work in the fuzzball arena has been on developing the supergravity side of the story and perturbative corrections to that picture. The CFT side of the story is less well developed, and it is this side to which we contribute here. 
Other recent CFT developments in fuzzball physics include using worldsheet CFT to perturbatively construct new microstate geometries for a class of D-brane boundstates known as superstrata \cite{Giusto:2012jx} and using D1-D5 CFT to perform spectral flow of integer and fractional type on the Ramond vacua of the twist sector and find microstate geometries corresponding to the double-centre Bena-Warner geometries \cite{Bena:2004de,Mathur:2011gz,Mathur:2012tj,Lunin:2012gp,Giusto:2012yz}.

From the microscopic string perspective, the degrees of freedom of the D1-D5 system are described by a conformal field theory (CFT) living on a symmetric product orbifold space. For D1-branes and D5-branes wrapped on $S^1$ and $S^1 \times T^4$ respectively, the orbifold is\footnote{Note that we are modding out by the symmetric group here, not the cyclic group.} $(T^4)^N/S_N$, where $N=N_1 N_5$.  Since our eventual goal is to help connect CFT physics with black hole physics, we are interested in what happens when the D1-D5 CFT is deformed {\em away} from the orbifold point. Throughout, we work in the large-$N$ limit where genus-zero diagrams dominate the string path integral.

In Section \ref{sectD1D5cftpert} we describe our method for identifying operator mixing among low-lying string states in the D1-D5 CFT. We outline the ingredients of the CFT, identify a suitable set of cocycles for the fermionic operators, and describe how conformal perturbation theory gives anomalous dimensions. We pick a deformation operator belonging to the twist-2 sector of the theory which is a singlet under R-symmetry and under the internal $SU(2)$s corresponding to directions of the $T^4$. We then show how taking factorization limits of four-point functions involving low-lying string states and the deformation operator allows us to identify operator mixing and to calculate mixing coefficients.

In Section \ref{dilaton}, we see why the dilaton operator does not mix, at first order in perturbation theory. This makes use of Lunin-Mathur (LM) technology for symmetric orbifolds \cite{Lunin:2000yv,Lunin:2001pw} and our companion work \cite{companion} generalizing LM to the non-twist sector.
In particular, we identify a suitable map from the base space to the covering surface and compute four-point functions.
Of course, the lack of mixing we find is in accord with expectations from non-renormalization theorems \cite{Baggio:2012rr}.

In Section \ref{lift} we present our main results.  First, we settle on a low-lying string state of the form $\partial X \partial X {\bar{\partial}}X {\bar{\partial}}X$. Then we discuss the technicalities of lifting the correlation function computation up to the covering space and summing over images. We next take the coincidence limit of the four-point function and show how to subtract conformal families of descendants in order to find the coefficient of mixing with other (quasiprimary) operators of
suitable weights. Finally, we evaluate the precise mixing coefficient for our string state. Our most interesting qualitative result is that mixing does take place at first order.

In section \ref{discsect}, we give a brief accessible summary of our results and discuss what remains to be done in order to nail down the anomalous dimensions for low-lying string states of interest.

\section{Perturbing the D1-D5 SCFT}\label{sectD1D5cftpert}

\subsection{The D1-D5 superconformal field theory}\label{cft}
The D1-D5 system is constructed by compactifying the type IIB string theory on $S^1\times \mathcal{M}^4$, where $\mathcal{M}^4$ is either $T^4$ or $K3$.
The bound state of $N_5$ D5 branes wrapping $S^1\times \mathcal{M}^4$ and $N_1$ D1 branes  wrapping $S^1$ defines the D1-D5 system.
In this work we consider the compactification on $T^4$.
If the size of the circle is much larger than the size of the torus, then in the low energy limit the system is described by a two-dimensional conformal field theory (CFT) living on $S^1$ \cite{David:2002wn}.
It is conjectured that there exists a point in the moduli space of the D1-D5 CFT at which the theory is a (1+1)-dimensional sigma model with target space being the symmetric product of $N_1\,N_5$ copies of $T^4$: $(T^4)^{N_1\,N_5}/S_{N_1\,N_5}$
\cite{Seiberg:1999xz,Vafa:1995bm,Larsen:1999uk,de Boer:1998ip,Dijkgraaf:1998gf,Arutyunov:1997gt,Arutyunov:1997gi,Jevicki:1998bm,Pakman:2009mi,Pakman:2009ab,Pakman:2009zz}.

The theory under consideration has a $\mathcal{N}=(4,4)$ supersymmetry.
It contains $SU(2)_L\times SU(2)_R$ R-symmetry and a $SO(4)_I\simeq SU(2)_1\times SU(2)_2$ symmetry group which corresponds to the directions of the torus.
(Technically the $SO(4)_I$ is broken by periodic identifications of the $T^4$, but it still provides a useful organizational principle.)
Each copy of the target space is a free $c=6$ CFT which has 4 real bosonic fields $X^1,\,X^2,\,X^3,\,X^4$, 4 real fermionic fields in the left moving sector $\psi^1,\,\psi^2,\,\psi^3,\,\psi^4$ and 4 real fermionic fields in the right moving sector $\tilde\psi^1,\,\tilde \psi^2,\,\tilde \psi^3,\,\tilde \psi^4$.

The bosonic fields $X^i$  can be written as doublets of $SU(2)_1$ and $SU(2)_2$ \cite{Avery:2009tu}
\begin{equation}\label{bosons}
X^{\dot AA}=\frac{1}{\sqrt2}\,X^i\,(\sigma^i)^{\dot AA},
\end{equation}
where $\sigma^1$, $\sigma^2$, and $\sigma^3$ are the Pauli spin matrices and $\sigma^4=i\,\mathrm{I}$, with $\mathrm{I}$ being the identity matrix.
The indices $A$ and $\dot A$ correspond to the doublets of $SU(2)_1$ and $SU(2)_2$, respectively.
The four real fermions of the left moving part can be combined to form complex fermions which transform as doublets of the $SU(2)_L$ and $SU(2)_2$: $\psi^{\alpha\dot A}$.
The index $\alpha$ corresponds to the $SU(2)_L$ doublet.
The reality condition imposes the following constraint on the complex fermions
\begin{equation}\label{ferl1}
(\psi^{\alpha\dot A})^\dagger(z)=
-\epsilon_{\alpha\beta}\,\epsilon_{\dot A\dot B}\,\psi^{\beta\dot B}(z).
\end{equation}
Thus we count four complex fermions $\psi^{\alpha\dot A}$, with two linearly independent reality conditions.  This gives two independent complex fermions, which is the same as four real fermions.
We similarly combine the four right moving real fermions into complex fermions which transform as doublets of the $SU(2)_R$ and $SU(2)_2$: $\tilde\psi^{\dot\alpha\dot A}$, where the index $\dot\alpha$ corresponds to the $SU(2)_R$ doublet.  Again, the reality condition \begin{equation}\label{ferr1}
(\tilde\psi^{\dot\alpha\dot A})^\dagger(\bar z)=-\epsilon_{\dot\alpha\dot\beta}\,\epsilon_{\dot A\dot B}\,\tilde\psi^{\dot\beta\dot B}(\bar z)
\end{equation}
reduces the counting to two independent complex fermions, or four real fermions.

The generators of the superconformal algebra in the left moving sector are the stress energy tensor $T$, the four supercurrents $G^{\alpha A}$, and the $SU(2)_L$ $R$-symmetry current $J^a$ which are given by
\begin{eqnarray}\label{gens}
T&=&\frac14\epsilon_{\dot A\dot B}\,\epsilon_{AB}\,
\partial X^{\dot AA}\,\partial X^{\dot BB}+
\frac12\epsilon_{\alpha\beta}\,\epsilon_{\dot A\dot B}\,
\psi^{\alpha\dot A}\partial\psi^{\beta\dot B},\nonumber\\
G^{\alpha A}&=&\sqrt2\epsilon_{\dot A\dot B}\,\psi^{\alpha\dot A}\,\partial X^{\dot BA},\nonumber\\
J^{a}&=&\frac14\epsilon_{\dot A\dot B}\,\psi^{\alpha\dot A}\,
(\sigma^{*a})^\beta_\gamma\,\psi^{\gamma\dot B}.
\end{eqnarray}
The generators of the superconformal algebra in the right moving sector are $\tilde T$, $\tilde G^{\dot\alpha A}$, and $\tilde J^a$ and are given by similar expressions.

The OPE of bosonic and fermionic fields are
\begin{eqnarray}\label{ope-x-l}
\partial X^{\dot AA}(z_1)\,\partial X^{\dot BB}(z_2)&\sim&
\frac{\epsilon^{\dot A\dot B}\,\epsilon^{AB}}{(z_1-z_2)^2},\\
\psi^{\alpha\dot A}(z_1)\,\psi^{\beta\dot B}(z_2)&\sim&
-\frac{\epsilon^{\alpha\beta}\,\epsilon^{\dot A\dot B}}{z_1-z_2}.
\end{eqnarray}
The left-moving generators of the CFT form a closed OPE current algebra, and similarly for the right-movers. The complete OPE algebra and the mode algebra may be found in \cite{Avery:2009tu}.

We introduce $N$ copies of the above CFT, and orbifold by the symmetric group $S_N$ which interchanges different copies of the CFT.  The orbifolding introduces twist sector states, and so in the CFT we have corresponding twist sector operators.  These twist operators mix the different copies of the CFT.
As one circles once around a point with a twist operator insertion $\sigma_n\equiv\sigma_{(123\cdots n)}$, the $n$ copies of the bosonic and fermionic fields are mapped together such that
\begin{eqnarray}\label{twist}
X^i_{(1)}\to X^i_{(2)}\to X^i_{(3)}\cdots\to X^i_{(n)}\to X^i_{(1)},\\
\psi^i_{(1)}\to\psi^i_{(2)}\to\psi^i_{(3)}\to\cdots\to\psi^i_{(n)}\to\psi^i_{(1)}.
\end{eqnarray}
This action is by itself not $S_N$ invariant, because it singles out the first $n$ indices.  One must in fact sum over the group orbit of the group element $(123\cdots n)$ in the full $S_N$ group, or equivalently to introduce one operator for each member of the conjugacy class of $(123\cdots n)$.  In what follows, we will consider only one member of the conjugacy class as a representative.  The contributions from other members of the conjugacy class will be accounted for with combinatoric factors later on.

An important property of the twist sector is that they contain operators with fractional modes.
The construction of these operators is explicitly described in \cite{Lunin:2001pw}.
Let us consider, for example, the $R$-current $J^a$ acting on the twist operator $\sigma_n$.
The fractional mode operator is defined
\begin{equation}\label{fracmodeJ}
J^a_{-\frac mn}(z)\,\sigma_n(0)\equiv\oint_{z=0}\frac{dz}{2\pi i}
\sum_{k=1}^{n}J^a_{(k)}(z)\,z^{-\frac mn}\,e^{-2\pi i\,(k-1)\frac mn}\,\sigma_n(0),
\end{equation}
where the subscript $(k)$ corresponds to the $k$-th copy of the target space involved in the twist.
The fractional mode operator is invariant under the action of the symmetric group because the integrand is periodic as one circles around the origin in the contour integral.
The operator $J^a_{-m/n}$ has a fractional holomorphic conformal weight $h=m/n$.

The fractional modes of the $R$-current (\ref{fracmodeJ}) are used to construct super chiral and anti-chiral primary operators of the twist sector.  This is done by starting with a bare twist, and then applying fractional R-current $J^+$ modes to find a superconformal ancestor \cite{Lunin:2001pw}.
Super chiral primaries have the same conformal weight and $R$-charge $h=m$, $\tilde h=\tilde m$ while super anti-chiral primaries satisfy $h=-m$, $\tilde h=-\tilde m$.

\subsection{Cocycles}\label{coc}
In the computations presented in the following sections we will frequently use the bosonized representation of the fermionic fields introduced in \cite{Lunin:2001pw}.
The bosonized language allows one to easily evaluate correlation functions which contain fermionic fields.
When bosonizing multiple (complex) fermions, one must introduce multiple bosonic fields.  These unrelated bosonic fields do not share an OPE, and so they commute as operators, while the fermions that they represent must anticommute.
This is a well known problem, and the introduction of cocycle operators is needed to guarantee that fermions anticommute.

Here, we will write down an explicit set of cocycles that guarantees the anticommutation of bosonized fermions in the D1-D5 CFT at the orbifold point.
In the bosonized language, we have two holomorphic scalar fields, $\phi_5, \phi_6$, and two antiholomorphic scalar fields, $\ti{\phi}_5, \ti{\phi}_6$ \footnote{The indices here are chosen to agree with those used in section 4 of \cite{Lunin:2001pw}.
We make no distinction between the index being up or down.}.
These fields have the corresponding momenta $p_5, p_6, \ti{p}_5,\ti{p}_6$ which satisfy the algebra
\be\label{qpcom}
[\phi_i,p_j]=i\delta_{ij}, \qquad [\ti \phi_i,\ti p_j]=i\delta_{ij},
\ee
where any of the $p_i$ and $\ti p_i$ commute, and the holomorphic and antiholomorphic sectors commute.

Fermions are replaced by exponentials of bosonic fields in the bosonized language.  Therefore, it is natural to consider operators of the form
\begin{equation}\label{op}
C_k\,e^{i\,(k_5\phi_5+k_6\phi_6+\tilde k_5\tilde\phi_5+\tilde k_6\tilde\phi_6)},
\end{equation}
where $k_i$ and $\ti k_j$ are real numbers and $C_k$ is the cocycle operator which explicitly depends on $k_i$ and $\ti k_j$.
We define the cocycle $C_k$ to be of the form
\begin{equation}\label{coc-gen}
C_k=e^{i\pi c_k},\qquad
c_k\equiv\begin{pmatrix} k_5 & k_6 & \tilde{k}_5 & \tilde{k}_6\end{pmatrix}
M \begin{pmatrix} p_5\\p_6\\\tilde p_5\\\tilde p_6\end{pmatrix},
\end{equation}
where $M$ is a $4\times4$ matrix.
Individual fermions are special operators for which one of the $k_i$ or $\ti k_i$ is 1, while the rest of them are 0.
Using this, we may constrain the matrix $M$ by requiring fermions to anticommute.  This gives that $M_{ij}-M_{ji}=\pm 1$ for $i\neq j$.
Thus, the antisymmetric part of $M$ is fixed, up to the choice of six signs, and the symmetric part is undetermined from such considerations.
We define the matrix $M$ to be of the form
\begin{equation}
M=\begin{pmatrix}\label{M}
\;\;\;\frac12 & \;\;\;\frac12 & -\frac12 & \;\;\;\frac12 \\
-\frac12 & -\frac12 & \;\;\frac12 & -\frac12 \\
\;\;\;\frac12 & -\frac12 & -\frac12 & -\frac12 \\
-\frac12 & \;\;\;\frac12 & \;\;\;\frac12 & \;\;\;\frac12
\end{pmatrix}.
\end{equation}
We choose this so that operators of the form $e^{\pm i \half (\phi_5+\phi_6)}$
or $e^{\pm i\half ({\tilde\phi_5} +{\tilde{\phi_6}})}$ are not dressed with any extra operators.

\subsubsection{Bosonizing the fermions}
We now explicitly write the cocycles associated with the four complex fermions in the holomorphic sector, $\psi^{\alpha \dot{A}}(z)$, and the four complex fermions in the anti-holomorphic sector, $\ti{\psi}^{\dot{\alpha}\dot{A}}(\bar z)$.
We impose the reality conditions (\ref{ferl1}) and  (\ref{ferr1}) to reproduce the correct number of two complex, or four real fermions in each sector.
When applying the reality conditions in the bosonized language, we must be careful to reverse the order of the exponentiated scalar field and the cocycle when we take the Hermitian conjugate
\be
(C_k e^{i k\cdot \phi})^{\dagger}=e^{-i k\cdot \phi} (C_k)^{\dagger}=e^{i (-k)\cdot \phi} C_{-k}.
\ee
Using the matrix $M$ defined above, we find
\bea\label{coc-fer-l}
 && \psi^{+ \dot{1}}=e^{i\pi c}\,e^{-i\phi_6}, \qquad\qquad
\psi^{+ \dot{2}}=e^{i\pi c}\,e^{i\phi_5},\nonumber\\
&& \psi^{-\dot{2}}=-e^{i\phi_6}\,e^{-i\pi c},\qquad\quad
\psi^{-\dot{1}}=e^{-i\phi_5} \,e^{-i\pi c},
\eea
where
\begin{equation}\label{coc-c}
e^{i\pi c}\equiv e^{\frac{i\pi}{2}\left(p_5 +p_6-\ti p_5+\ti p_6\right)}.
\end{equation}
One can move all the cocycle operators to the left by using the commutation relations (\ref{qpcom}), if so desired.
Using similar considerations for the antiholomorphic side, we find
\bea\label{coc-fer-r}
&&\ti\psi^{\dot+ \dot{1}}=e^{i\pi\tilde c}\,e^{-i\ti\phi^6}, \qquad\qquad
\ti\psi^{\dot+ \dot{2}}=e^{i\pi\tilde c}\,e^{i\ti\phi^5},\nonumber\\
&& \ti\psi^{\dot- \dot{2}}=-e^{i\ti\phi^6}\,e^{-i\pi\tilde c}, \qquad\quad
\ti\psi^{\dot- \dot{1}}=e^{-i\ti\phi^5}\,e^{-i\pi\tilde c},
\eea
where
\begin{equation}\label{coc-tildec}
e^{i\pi\tilde c}\equiv e^{\frac{i\pi}{2}\left(p_5 -p_6-\ti p_5-\ti p_6\right)}.
\end{equation}

One can easily check that the assigned cocycles guarantee the anti-commutation of fermions.
Let us first consider bosonized fermions which contain singular terms in their OPEs.
The singular part is proportional to $(z_1-z_2)^{-1}$, where $z_1$ and $z_2$ are the positions of the fermion operators on the complex plane.
Therefore, a minus sign arises when we switch the order of the two operators.
For those bosonized fermions which do not share a singular OPE, passing the cocyle operators across the exponentiated scalars produces the minus sign.
For instance, we consider the OPE
\begin{eqnarray}\label{ex-fer-fer}
\psi^{+\dot1}(z)\,\psi^{+\dot2}(w)&=&
e^{i\frac{\pi}{2}(p_5+p_6-\tilde p_5+\tilde p_6)}e^{-i\phi_6}(z)\,
e^{i\frac{\pi}{2}(p_5+p_6-\tilde p_5+\tilde p_6)}e^{i\phi_5}(w)\nonumber\\
&=&e^{[-i\phi_6,i\frac{\pi}{2}p_6]}\,e^{i\frac{\pi}{2}(p_5+p_6-\tilde p_5+\tilde p_6)}\,
e^{i\frac{\pi}{2}(p_5+p_6-\tilde p_5+\tilde p_6)}e^{-i\phi_6}(z)\,e^{i\phi_5}(w)\nonumber\\
&=&e^{[-i\phi_6,i\frac{\pi}{2}p_6]}\,e^{[i\frac{\pi}{2}p_5,i\phi_5]}\,
e^{i\frac{\pi}{2}(p_5+p_6-\tilde p_5+\tilde p_6)}e^{i\phi_5}(w)\,
e^{i\frac{\pi}{2}(p_5+p_6-\tilde p_5+\tilde p_6)}e^{-i\phi_6}(z)\nonumber\\
&=&-\psi^{+\dot2}(w)\,\psi^{+\dot1}(z),
\end{eqnarray}
where we have used the Baker-Campbell-Hausdorff formula and the commutation relations (\ref{qpcom}).

We noted earlier that the fermions in the holomorphic sector transform as doublets of $SU(2)_L$ and $SU(2)_2$ and that the fermions in the anti-holomorphic sector transform as doublets of $SU(2)_R$ and $SU(2)_2$.
Using the prescription described above for assigning cocyles to the bosonized fields, one can construct the $SU(2)$ symmetry currents.
Let us consider the holomorphic sector first.
Equation (\ref{gens}) gives the $R$-currents $J^+=\psi^{+\dot1}\psi^{+\dot2}$ and $J^-=-\psi^{-\dot1}\psi^{-\dot2}$.
Using (\ref{coc-fer-l}) and (\ref{coc-c}), we can then evaluate the OPE of the $R$-currents $J^{\pm}$ and the fermions in the context of the bosonized language.
Let us evaluate the OPEs for $\psi^{\pm\dot1}$ as an example.
For $\psi^{\pm\dot1}$ we have:
\begin{eqnarray}\label{ex-J-fer-1}
J^{+}(z)\psi^{-\dot1}(w)&=&
e^{i\frac{\pi}{2}(p_5+p_6-\tilde p_5+\tilde p_6)}e^{-i\phi_6}
e^{i\frac{\pi}{2}(p_5+p_6-\tilde p_5+\tilde p_6)}e^{i\phi_5}(z)\;\;
e^{-i\phi_5}(w)\,e^{-i\frac{\pi}{2}(p_5+p_6-\tilde p_5+\tilde p_6)}\nonumber\\
&\sim&\frac{e^{i\frac{\pi}{2}(p_5+p_6-\tilde p_5+\tilde p_6)}e^{-i\phi_6}(w)}{z-w}\nonumber\\
&=&\frac{\psi^{+\dot1}(w)}{z-w},
\end{eqnarray}
and
\begin{eqnarray}\label{ex-J-fer-2}
J^{-}(z)\psi^{+\dot1}(w)&=&
e^{-i\phi_5}\,e^{-i\frac{\pi}{2}(p_5+p_6-\tilde p_5+\tilde p_6)}
e^{i\phi_6}e^{-i\frac{\pi}{2}(p_5+p_6-\tilde p_5+\tilde p_6)}(z)\;\;
e^{i\frac{\pi}{2}(p_5+p_6-\tilde p_5+\tilde p_6)}e^{-i\phi_6}(w)\nonumber\\
&\sim&\frac{e^{-i\phi_5}(w)\,e^{-i\frac{\pi}{2}(p_5+p_6-\tilde p_5+\tilde p_6)}}
{z-w}\nonumber\\
&=&\frac{\psi^{-\dot1}(w)}{z-w}.
\end{eqnarray}
Similar considerations hold for $\psi^{\pm\dot2}$.
We thus obtain
\begin{equation}\label{Jpsi}
[J^+_0,\psi^{-\dot A}]=\psi^{+\dot A},\qquad[J^-_0, \psi^{+\dot A}]=\psi^{-\dot A}.
\end{equation}
Therefore the bosonized fermions transform as a doublet of the $SU(2)_L$ symmetry, as required.
One can perform similar computations in the right moving sector and find that the assigned cocycles reproduce the appropriate $SU(2)_R$ algebra.

One can also easily check that the fermions transform as doublets of $SU(2)_2$ in the context of the bosonized language.
Let us denote the associated currents with  $\mathcal J^a$.
We have
$\mathcal J^+=\psi^{+\dot1}\,\psi^{-\dot1},\quad
\mathcal J^-=-\psi^{+\dot2}\,\psi^{-\dot2}$.
Using the bosonized fields (\ref{coc-fer-l}) and (\ref{coc-fer-r}) and performing similar computations as in (\ref{ex-J-fer-1}) and (\ref{ex-J-fer-2}), we find
\begin{equation}\label{mathcalJpsi}
[\mathcal J^+_0,\psi^{\alpha\dot2}]=\psi^{\alpha\dot1},
\qquad[\mathcal J^-_0, \psi^{\alpha\dot1}]=\psi^{\alpha\dot2},
\end{equation}
Analogous considerations hold in the right moving sector.

\subsubsection{Spin fields}
Our next task is to figure out how to make spin fields work in the Lunin-Mathur context.
The main ingredients are as follows; we refer the interested reader to \cite{Lunin:2001pw} for further details.
First, we use Lunin-Mathur technology to map twist sector operators into the covering space where they have normal boundary conditions. Second, fermions belonging to the even-twist sector have antiperiodic boundary conditions in the cover. Third, to take fermion boundary conditions into account correctly, spin fields must be inserted in the cover.

We now turn to constructing the cocycles associated with the spin fields on the covering surface.
For spin fields we have $k_i=\pm1/2$, $\ti k_i=\pm 1/2$.
Let us consider the holomorphic sector first.
There are four spin fields in this sector $\mathcal{S}^{\alpha}$ and $\mathcal{S}^{\dot A}$.
Following the procedure described above, we find that
\be
{\mathcal S}^+\equiv e^{i\pi c}\,e^{\frac{i}{2}\left(\phi_5-\phi_6\right)}, \qquad\qquad
{\mathcal S}^-\equiv  e^{-i\pi c}\,e^{-\frac{i}{2}\left(\phi_5-\phi_6\right)},
\ee
where $e^{i\pi c}$ is defined in (\ref{coc-c}).
In the above expressions we have moved the cocycles to the left.
For these spin fields we find that
\be
[J^+_0, {\mathcal S}^-]={\mathcal S}^+,\qquad[J^-_0, {\mathcal S}^+]={\mathcal S}^-,
\qquad ({\mathcal S}^+)^\dagger={\mathcal S}^-.
\ee
The assigned cocycles make the two spin fields $\mathcal{S}^{\alpha}$ transform as a doublet of $SU(2)_L$ and a singlet of $SU(2)_2$.

For spin fields $\mathcal{S}^{\dot A}$ we find
\be
{\mathcal S}^{\dot1}\equiv e^{i\frac{\pi}{2}}\,e^{-\frac{i}{2}\left(\phi_5+\phi_6\right)},
\qquad\qquad
{\mathcal S}^{\dot2}\equiv  e^{-i\frac{\pi}{2}}\,e^{\frac{i}{2}\left(\phi_5+\phi_6\right)}.
\ee
We note that the structure of the cocycles we defined in (\ref{coc-gen}) and (\ref{M}) is such that the cocycles associated with exponentials of the form $e^{\pm\frac{i}{2}(\phi_5+\phi_6)}$ are just 1.
The rescaling of the spin fields with the above phases guarantees that they satisfy the $SU(2)_2$ algebra
 \be
[\mathcal{J}^+_0, {\mathcal S}^{\dot2}]={\mathcal S}^{\dot1},\qquad
[\mathcal{J}^-_0, {\mathcal S}^{\dot2}]={\mathcal S}^{\dot2},
\qquad ({\mathcal S}^{\dot1})^\dagger={\mathcal S}^{\dot2}.
\ee
These two spin fields do not carry $SU(2)_L$ charges.

Analogous computations are done for the anti-holomorphic sector and we find that the spin fields in this sector are given by
\bea
&&\mathcal{\tilde S}^{\dot+}\equiv
e^{i\frac{\pi}{2}(p_5-p_6-\tilde p_5-\tilde p_6)}e^{\frac{i}{2}\left(\ti \phi_5-\ti \phi_6\right)}, \qquad
\mathcal{\tilde S}^{\dot-}\equiv
e^{-i\frac{\pi}{2}(p_5-p_6-\tilde p_5-\tilde p_6)}e^{-\frac{i}{2}\left(\ti \phi_5-\ti \phi_6\right)},\\
&&{\tilde{\mathcal S}}^{\dot1}\equiv
e^{-i\frac{\pi}{2}}\,e^{-\frac{i}{2}\left(\tilde\phi_5+\tilde\phi_6\right)},
\qquad\qquad\qquad
{\tilde{\mathcal S}}^{\dot2}\equiv
e^{i\frac{\pi}{2}}\,e^{\frac{i}{2}\left(\tilde\phi_5+\tilde\phi_6\right)}.
\eea

Finally, we note the the fermionic zero modes act on the spin fields and map them to other spin fields.
It is straightforward to check that these modes satisfy the gamma matrix algebra  $\{\Gamma^i,\Gamma^j\}=2\,\delta^{ij}$.
The action of fermion zero modes on the spin fields can be evaluated using the bosonized language. For this purpose, and some later ones, we now record the OPEs of the fermions and the spin fields.
In the left moving sector we find that
\begin{eqnarray}\label{psi0S-l}
&&\psi^{-\dot1}_0\,\mathcal{S}^+(0)=
e^{-i\frac{\pi}{2}}\,\mathcal{S}^{\dot1}(0),
\qquad\qquad\qquad
\psi^{-\dot2}_0\,\mathcal{S}^{+}(0)=
e^{-i\frac{\pi}{2}}\,\mathcal{S}^{\dot2}(0),\label{S+}\nonumber\\
&&\psi^{+\dot1}_0\,\mathcal{S}^-(0)=
-\mathcal{S}^{\dot1}(0),
\qquad\qquad\qquad\quad\;
\psi^{+\dot2}_0\,\mathcal{S}^-(0)=
-\mathcal{S}^{\dot2}(0),\nonumber\\
&&\psi^{+\dot2}_0\,\mathcal{S}^{\dot1}(0)=
e^{+i\frac{\pi}{2}}\,\mathcal{S}^{+}(0),\qquad\qquad\quad\;\;\;\,
\psi^{-\dot2}_0\,\mathcal{S}^{\dot1}(0)=
\mathcal{S}^{-}(0),\nonumber\\
&&\psi^{+\dot1}_0\,\mathcal{S}^{\dot2}(0)=
e^{-i\frac{\pi}{2}}\,\mathcal{S}^{+}(0),\qquad\qquad\quad\;\;\;\,
\psi^{-\dot1}_0\,\mathcal{S}^{\dot2}(0)=
-\mathcal{S}^{-}(0).
\end{eqnarray}
Similar considerations apply for the right-moving sector.

\subsection{Deformation operator}\label{dfmn}
The  D1-D5 CFT at the orbifold point has four exact marginal deformation
operators in the twist sector which deform the CFT away from the orbifold point.
These deformation operators have conformal weights $(h,\bar{h})=(1,1)$.
They are singlets under the $SU(2)_L\times SU(2)_R$ $R$-symmetry and preserve the $\mathcal{N}=(4,4)$ supersymmetry of the CFT. 
The deformation operators are constructed by applying modes of supercurrent
$G_{-1/2}^{\alpha A}$ and $\tilde{G}_{-1/2}^{\dot\alpha B}$ to the super-chiral and
anti-chiral primaries of the twist-2 sector, $\sigma_2^{\beta\dot\beta}$, with conformal
dimension $(1/2,1/2)$.
The left and right moving parts of the four marginal deformation operators carry
indices of the doublet of $SU(2)_1$ of the internal $SO_I(4)$ symmetry and the
operators transform as $\bf{3}+\bf{1}$ of $SU(2)_1$ \cite{David:2002wn}.

We will consider the singlet component which preserves the global $SU(2)_1$
symmetry:
\begin{equation}\label{odl}
\mathcal{O}_d\propto\epsilon_{AB}\,\epsilon_{\alpha\beta}\,\epsilon_{\dot\alpha\dot\beta}\,
G_{-\frac12}^{\alpha A}\,\tilde G_{-\frac12}^{\dot\alpha B}\,\sigma_2^{\beta\dot\beta}.
\end{equation}
We now split the above twist operator into left-moving and right-moving parts $\sigma_2^{\beta\dot\beta}(z,\zb)=\sigma_2^{\beta}(z) \sigma_2^{\dot \beta}(z)$.  It is shown in \cite{Avery:2010er} (Appendix B) that
$G_{-\frac12}^{-A}\,\sigma_2^{+}\propto G_{-\frac12}^{+A}\,\sigma_2^{-}$.
Therefore, $G_{-\frac12}^{-A}\,\sigma_2^{+}$ is a singlet of $SU(2)_L$ by itself.
The same reasoning holds for the right moving sector.
This could also be seen by the discussion in \cite{Avery:2010qw} (\S5.1) that a
single application of supercharges $G_{-\frac12}^{-A}$ on a super-chiral primary
$\sigma_n^+$ constructs Virasoro primaries which are annihilated by $J_0^+$ and
therefore are the top members of the $SU(2)_L$ multiplet.
In the twist-2 sector this construction creates a singlet under $SU(2)_L$.
The deformation operator with charge zero under $SU(2)_L\times SU(2)_R$ and
$SU(2)_1$ is therefore of the form:
\begin{equation}\label{ods1}
\mathcal{O}_d=\epsilon_{AB}\,G_{-\frac12}^{-A}\,\tilde G_{-\frac12}^{\dot-B}\,
\sigma_2^{+\dot+}.
\end{equation}
The perturbation that we add to deform the symmetric product CFT away from the
orbifold point is
\begin{equation}
S_{\mathrm{int}}=\lambda \int d^2z\,\mathcal{O}_d(z,\bar z)+\mathrm{a.c.},
\end{equation}
where a.c. refers to the anti-chiral fields acted on by the Hermitian conjugate
supercurrent modes.

The supercurrents are
$G^{\alpha A}=\sqrt2\,\psi^{\alpha \dot A}\partial X^{\dot BA}\epsilon_{\dot A\dot B}$,
where the summation over an omitted target space copy index is implicit.
Inserting this in (\ref{ods1}) we obtain:
\begin{eqnarray}\label{ods2}
\mathcal{O}_d&=&\sqrt2\,\Big[
(\psi^{-\dot 1}\partial X^{\dot 21}-\psi^{-\dot 2}\partial X^{\dot11})_{-\frac12}\,
(\tilde\psi^{\dot-\dot 1}\bar\partial X^{\dot 22}-
\tilde\psi^{\dot-\dot 2}\bar\partial X^{\dot12})_{-\frac12}\nonumber\\
&&\quad-(\psi^{-\dot 1}\partial X^{\dot 22}-\psi^{-\dot 2}\partial X^{\dot12})_{-\frac12}\,
(\tilde\psi^{\dot-\dot 1}\bar\partial X^{\dot 21}-
\tilde\psi^{\dot-\dot 2}\bar\partial X^{\dot11})_{-\frac12}
\Big]\sigma_2^{+\dot+}.
\end{eqnarray}
The modes of the supercurrents in the left moving sector are given by
\begin{equation}\label{sucum}
G^{\alpha A}_{m}=\oint \frac{dz}{2\pi i}\,G^{\alpha A}\,z^{h+m-1},
\end{equation}
where $h=3/2$.
The same procedure holds for the right moving supercurrents
$\tilde G^{\dot\alpha A}_{m}$.

In the twist-2 sector, the insertion of super (anti-)chiral primary $\sigma_2^{\pm\dot\pm}$ operators in the base space corresponds to the insertion of spin fields $\mathcal{S}^{\pm\dot\pm}$ in the covering surface \cite{Lunin:2001pw}.
The operators
$\sigma_2^{\pm\dot\pm}$ that we are concerned with have conformal weight
$(1/2,1/2)$ and carry $R$-charge $(\pm1/2,\pm1/2)$.
We discussed the bosonized representation of fermions and spin fields in subsection \ref{coc}.
Here we will use them to evaluate the deformation operator.
It is shown in \cite{Lunin:2001pw} that the local normalization of the spin fields is given by $b^{-\frac18}\,\mathcal{S}^{\pm}$ where $b$ is specified by the map from the base space to the cover in the vicinity of the insertion point of the spin field: $(z-z_0)\approx b\,(t-t_0)^2\,(1+b_1t+O(t^2))$.
The right moving part definitions follows similarly. 
In cases where there are more than two spin fields or there are fermionic fields acting
on the spin fields in a correlation function, one needs to be careful about the anticommutation relations of fermionic fields and apply cocycles to impose the correct anticommutation properties.
This is the raison d'\^{e}tre for subsection \ref{coc}.

We will now construct $\mathcal{O}_d$.
Using (\ref{sucum}) we have:
\begin{eqnarray}\label{ggi}
&& G_{-\frac12}^{-A}\,\tilde G_{-\frac12}^{\dot-B}\,\sigma_2^{+\dot+}(0,0)=
\oint_{0}\frac{dz}{2\pi i}\oint_{0}\frac{d\bar z}{2\pi i}\,
z^{\frac32-\frac12-1}\;\bar z^{\frac32-\frac12-1}
\,G^{-A}(z)\,\tilde G^{\dot-B}(\bar z)\,\sigma_2^{+\dot +}(0,0)
\nonumber\\
&&\quad\to\oint_{0}\frac{dt}{2\pi i}\left(\frac{dz}{dt} \right)^{1-\frac{3}{2}}
\oint_{0}\frac{d\bar t}{2\pi i}\left(\frac{d\bar z}{d\bar t} \right)^{1-\frac{3}{2}}
G^{-A}(t)\,\tilde G^{\dot-B}(\bar t)\left(\frac{1}{b^{\frac18}\bar b^{\frac18}}
\mathcal{S}^{+\dot+}(0)\right)
\nonumber\\
&&\quad=\oint_{0}\frac{dt}{2\pi i}\oint_{0}\frac{d\bar t}{2\pi i}
\frac{1}{\Big(2bt(1+\frac32\frac{b_1}{b} t+O(t^2))\Big)^{\frac12}}
\frac{1}{\Big(2\bar b \bar t(1+\frac32\frac{\bar b_1}{\bar b} \bar t+O(\bar t^2))\Big)^{\frac12}}
\times
\nonumber\\
&&\qquad\quad \times\  G^{-A}(t)\,\tilde G^{\dot-B}(\bar t)
\frac{1}{|b|^{\frac14}}\mathcal{S}^{+\dot+}(0)
\nonumber\\
&&\quad\approx\frac{2}{2|b|^{\frac54}}\oint_{0}\frac{dt}{2\pi i}\oint_{0}\frac{d\bar t}{2\pi i}
\frac{\Big(1-\frac34\frac{b_1}{b}\,t+O(t^2)\Big)}{t^{\frac12}}
\frac{\Big(1-\frac34\frac{\bar b_1}{\bar b}\,t+O(\bar t^2)\Big)}{\bar t^{\frac12}}\times
\nonumber\\
&&\qquad\quad \times
\Big(\psi^{-\dot1}\partial X^{\dot2A}-\psi^{-\dot2}\partial X^{\dot1A}\Big)(t)\,
\Big(\tilde\psi^{\dot-\dot1}\bar\partial X^{\dot2B}-
\tilde\psi^{\dot-\dot2}\bar\partial X^{\dot1B}\Big)(\bar t)\,
\mathcal{S}^{+\dot+}(0,0),
\end{eqnarray}
where the arrow in the second line implies that we passed from the base to the
covering sphere using the map which is locally of the form
$z\approx b\,t^2\,(1+b_1t+O(t^2))$.
Here and in what follows, we simplify notation by suppressing the contribution from the Liouville action which is always present \cite{Lunin:2000yv,Lunin:2001pw}.  Thus $\rightarrow$ is really an instruction to pass to the covering surface, and suppress the Liouville contribution.  We will put in this contribution in at the end.
The OPEs of fermions and spin fields are evaluated in (\ref{psi0S-l}) in the context of the bosonized language.
The bosonic fields $\partial X^{\dot AA}$ do not share singular OPEs with the spin
fields.
Equation (\ref{ggi}) then reads:
\begin{eqnarray}\label{ggii}
&&G_{-\frac12}^{-A}\,\tilde G^{\dot-B}\,\sigma_2^{+\dot+}(0,0)=
\frac{1}{|b|^{\frac54}}\oint_{0}\frac{dt}{2\pi i}\oint_{0}\frac{d\bar t}{2\pi i}
\,\frac{1}{t}\,\frac{1}{\bar t}\times
\nonumber\\
&&\qquad\qquad\qquad\qquad
\Big(:\partial X^{\dot2A}\bar\partial X^{\dot2B}e^{-i\phi_5(t)-i\tilde\phi_5(\bar t)}\,
e^{\frac{i}{2}(-\phi_6+\phi_5-\tilde\phi_6+\tilde\phi_5)(0,0)}:\nonumber\\
&&\qquad\qquad\qquad\qquad
+:\partial X^{\dot2A}\bar\partial X^{\dot1B}e^{-i\phi_5(t)+i\tilde\phi_6(\bar t)}\,
e^{\frac{i}{2}(-\phi_6+\phi_5-\tilde\phi_6+\tilde\phi_5)(0,0)}:\nonumber\\
&&\qquad\qquad\qquad\qquad
+:\partial X^{\dot1A}\bar\partial X^{\dot2B}e^{+i\phi_6(t)-i\tilde\phi_5(\bar t)}\,
e^{\frac{i}{2}(-\phi_6+\phi_5-\tilde\phi_6+\tilde\phi_5)(0,0)}:\nonumber\\
&&\qquad\qquad\qquad\qquad
+:\partial X^{\dot1A}\bar\partial X^{\dot1B}e^{+i\phi_6(t)+i\tilde\phi_6(\bar t)}\,
e^{\frac{i}{2}(-\phi_6+\phi_5-\tilde\phi_6+\tilde\phi_5)(0,0)}:\Big)\nonumber\\
&=&\frac{1}{|b|^{\frac54}}\,
\Big(:\partial X^{\dot2A}\bar\partial X^{\dot2B}
e^{\frac{i}{2}(-\phi_6-\phi_5-\tilde\phi_6-\tilde\phi_5)}:
+:\partial X^{\dot2A}\bar\partial X^{\dot1B}
e^{\frac{i}{2}(-\phi_6-\phi_5+\tilde\phi_6+\tilde\phi_5)}:\nonumber\\
&&\qquad\!+:\partial X^{\dot1A}\bar\partial X^{\dot2B}
e^{\frac{i}{2}(+\phi_6+\phi_5-\tilde\phi_6-\tilde\phi_5)}:
+:\partial X^{\dot1A}\bar\partial X^{\dot1B}
e^{\frac{i}{2}(+\phi_6+\phi_5+\tilde\phi_6+\tilde\phi_5)}:\Big)(0,0),\nonumber\\
\end{eqnarray}
up to an overall minus sign coming from moving the right sector bosonized fermions
to the right to evaluate their OPEs with the right moving spin fields. 
The deformation operator (\ref{ods2}) is then found to be of the form
\begin{eqnarray}\label{defool}
\mathcal{O}_d\,(0,0)=\frac{1}{|b|^{\frac54}}&\bigg(&
:e^{\frac{i}{2}(-\phi_6-\phi_5-\tilde\phi_6-\tilde\phi_5)}
\left(\partial X^{\dot21}\bar\partial X^{\dot22}-\partial X^{\dot22}\bar\partial X^{\dot21}
\right):\nonumber\\
&+&:e^{\frac{i}{2}(-\phi_6-\phi_5+\tilde\phi_6+\tilde\phi_5)}
\left(\partial X^{\dot21}\bar\partial X^{\dot12}-\partial X^{\dot22}\bar\partial X^{\dot11}
\right):\nonumber\\
&+&:e^{\frac{i}{2}(+\phi_6+\phi_5-\tilde\phi_6-\tilde\phi_5)}
\left(\partial X^{\dot11}\bar\partial X^{\dot22}-\partial X^{\dot12}\bar\partial X^{\dot21}
\right):\nonumber\\
&+&:e^{\frac{i}{2}(+\phi_6+\phi_5+\tilde\phi_6+\tilde\phi_5)}
\left(\partial X^{\dot11}\bar\partial X^{\dot12}-\partial X^{\dot12}\bar\partial X^{\dot11}
\right):\bigg)(0,0).
\end{eqnarray}

The complete deformation is $\lambda\mathcal{O}_d+\mathrm{a.c.}$, where a.c. refers to the action of the Hermitian conjugated supercurrent modes on the super anti-chiral primary fields $\sigma_2^{-\dot-}$.
The super chiral and anti-chiral primary operators of the left and right moving sectors form doublets of $SU(2)_L$ and $SU(2)_R$, respectively.
As discussed in \cite{Gava:2002xb}, we can write the left moving super anti-chiral operator as
\begin{equation}\label{sigma2--}
\sigma_2^{-\dot-}(z_0,\bar z_0)=
J_0^-(z)\,\tilde J_0^{\dot-}(\bar z)\,\sigma_2^{+\dot+}(z_0,\bar z_0).
\end{equation}
The anti-chiral part of the deformation operator is of the form
\begin{equation}\label{odsc2}
G^{+A}_{-\frac12}(z^\prime)\,\tilde G^{\dot+B}_{-\frac12}(\bar z^\prime)\,\sigma_2^{-\dot-}=
G^{+A}_{-\frac12}(z^\prime)\,\tilde G^{\dot+B}_{-\frac12}(\bar z^\prime)\,
J_0^-(z)\,\tilde J_0^{\dot-}(\bar z)\,\sigma_2^{+\dot+}(z_0,\bar z_0).
\end{equation}
One can then pass $J^-$ and $\tilde J^{\dot-}$ to the left of $G^{+A}$ and $\tilde G^{\dot+A}$ which gives $G^{-A}$ and $\tilde G^{\dot-A}$ acting on $\sigma_2^{+\dot+}(z_0,\bar z_0)$.
Therefore the two chiral and anti-chiral terms are identical and the full deformation operator is of the form $\lambda\,\mathcal{O}_d$, where $\lambda$ is a real number. 
This can be checked explicitly in the context of the bosonized language noting that the spin fields $\mathcal{S}^{\pm}$ and $\mathcal{S}^{\dot{\pm}}$ carry $SU(2)_L$ and $SU(2)_R$ indices and taking into account $\epsilon_{\alpha\beta}$  and $\epsilon_{\dot\alpha\dot\beta}$ factors when taking their Hermitian conjugates. 
One may be able to account for these signs in a more explicit way with more complicated cocycles \cite{Kostelecky:1986xg}.

Lastly, we note that we work with $S_N$-invariant operators in the symmetric product orbifold, which are constructed by summing over the conjugacy classes.
The summation brings a combinatorial factor (which depends on $N$ and $n$) in the definition of the operators in the twist-$n$ sector.  We will discuss this later.

\subsection{Conformal perturbation theory}\label{cpft}

We consider a quasi-primary field $\phi_i$ in the 2-dimensional $\mathcal{N}=(4,4)$
D1-D5 SCFT.
The two-point function of $\phi_i$ and other quasi-primaries read:
\begin{equation}\label{2pf1}
\langle\phi^i(z_1,\bar z_1)\,\phi^{j}(z_2,\bar z_2)\rangle_0=
\frac{\delta^{ij}}{z_{12}^{h_i}\,\bar z_{12}^{\tilde h_i}},
\end{equation}
where the subscript ``0" refers to the unperturbed CFT, $h_i$ ($\tilde h_i$) are
the unperturbed (anti-) holomorphic conformal weights of $\phi_i$, and
$z_{12}\equiv z_1-z_2$.
Let us first assume that there is \emph{no} degeneracy in the conformal weight of
$\phi_i$, \emph{i.e.}, if $h_i=h_j$, then $\phi_i=\phi_j$.
We first perform the perturbation theory under this assumption and then generalize to the case where multiple fields can have the same conformal weight.

We start by adding a small perturbation to the action of the free SCFT:
\begin{equation}\label{pert}
\delta S=\sum_A\lambda_A\int d^2z \mathcal{O}_{d,A}(z,\bar z),
\end{equation}
where $\lambda_A$ are the coupling constants of perturbation and
$\mathcal{O}_{d,A}(z,\bar z)$ are the exact marginal operators of the theory.
We perform Kadanoff's conformal perturbation theory and evaluate the
deformation of the conformal weights \cite{kadanoff78i}.
Two-point correlation functions of quasi primary operators of a CFT are determined by their conformal weights. Thus, their deformation gives the deformation of the conformal weights.
There are two ways to evaluate the change in the two-point function.
First, one evaluates the derivative of the two-point function (\ref{2pf1}) with respect to
the coupling constant:
\begin{eqnarray}
\langle \phi^i(z_1,\bar z_1)~\phi^j(z_2,\bar{z_2})\rangle_{\lambda_A}
&=&\frac{\delta^{ij}}{z_{12}^{2h_i(\lambda_A)}
\bar z_{12}^{2\tilde{h_i}(\lambda_A)}}
=\frac{\delta^{ij}}
{z_{12}^{2\left(h_i+\lambda_A\frac{\partial h_i(\lambda_A)}{\partial\lambda_A}\right)}
\bar z_{12}^{2\left(\tilde{h}_i+\lambda_A\frac{\partial\tilde h_i(\lambda_A)}{\partial\lambda_A}\right)}}
\nonumber \\
&=&e^{-2\lambda_A\left(\frac{\partial h_i(\lambda_A)}{\partial\lambda_A}\ln(z_{12})
+\frac{\partial\tilde h_i(\lambda_A)}{\partial\lambda_A}\ln(\bar z_{12})\right)}
\langle \phi^{i}(z_1,\bar z_1)~\phi^{j}(z_2,\bar{z_2})\rangle_{0}\nonumber
\end{eqnarray}
\begin{equation}\label{2pfb1}
\approx
\left(1-2\lambda_A\left(\frac{\partial h_i(\lambda_A)}{\partial\lambda_A}\ln(z_{12})
+\frac{\partial\tilde h_i(\lambda_A)}{\partial\lambda_A}\ln(\bar z_{12})\right)\right)
\langle \phi^{i}(z_1,\bar z_1)\phi^{j}(z_2,\bar{z_2})\rangle_{0},
\end{equation}
where $h_i(\lambda_A)=h_i+\lambda_A\frac{\partial h_i(\lambda_A)}
{\partial\lambda_A}$ and $\tilde h_i(\lambda_A)=\tilde h_i+\lambda_A\frac
{\partial\tilde h_i(\lambda_A)}{\partial\lambda_A}$
to the first order in perturbation theory.
We then have:
\begin{eqnarray}\label{2der1i}
&& \frac{\partial}{\partial\lambda_A}
\langle \phi^i(z_1,\bar z_1)~\phi^j(z_2,\bar{z_2})\rangle_{\lambda_A} = \nonumber\\
&&\qquad\qquad\left(-2\frac{\partial h_i(\lambda_A)}{\partial\lambda_A}\ln(z_{12})
-2\frac{\partial\tilde h_i(\lambda_A)}{\partial\lambda_A}\ln(\bar z_{12})\right)
\langle \phi^i(z_1,z_2)~\phi^j(z_2,\bar{z_2})\rangle_{0}.
\end{eqnarray}
If the derivatives of the conformal weights $(h_i(\lambda_A),\tilde h_i(\lambda_A))$
vanish to the first order, then one moves to the second order in perturbation theory:
\begin{eqnarray}\label{2pf3}
&& \frac{\partial^2}{\partial\lambda_A^2}
\langle\phi^i(z_1,\bar z_1)\,\phi^{j}(z_2,\bar z_2)\rangle
_{\lambda_A} \nonumber\\
&&\qquad\qquad=\left(-2\frac{\partial^2 h_i(\lambda_A)}{\partial\lambda^2}\ln(z_{12})
-2\frac{\partial^2 \tilde h_i(\lambda_A)}{\partial\lambda^2}\ln(\bar z_{12}) \right)
\langle\phi_i(z_1,\bar z_1)\,\phi^{\dagger}_j(z_2,\bar z_2)\rangle_0,\qquad
\end{eqnarray}
and so on.

The second way of evaluating the change in the two-point function is to use the
path integral formulation of the theory:
\begin{eqnarray}
\langle\phi^i(z_1,\bar z_1)\,\phi^j(z_2,\bar z_2)\rangle_{\lambda_A}&=&
\frac{\int d[X,\psi]
\,e^{-S_{\mathrm{free}}+\lambda_A\int\,d^2z\mathcal{O}_{d,A}(z,\bar z)}\,
\phi^i(z_1,\bar z_1)\,\phi^j(z_2,\bar z_2)}
{\int d[X,\psi]\,e^{-S_{\mathrm{free}}+\lambda_A\int\,d^2z\mathcal{O}_{d,A}(z,\bar z)}}
\nonumber
\end{eqnarray}
\begin{equation}\label{2pf4}
=\frac{\int d[X,\psi]
\,e^{-S_{\mathrm{free}}}
\Big(1+\lambda_A\int\,d^2z\mathcal{O}_{d,A}(z,\bar z)+O(\lambda_A^2)\Big)
\,\phi^i(z_1,\bar z_1)\,\phi^j(z_2,\bar z_2)}
{\int d[X,\psi]\,e^{-S_{\mathrm{free}}}
\Big(1+\lambda_A\int\,d^2z\mathcal{O}_{d,A}(z,\bar z)+O(\lambda_A^2)\Big)},
\end{equation}
where we expanded the perturbative terms to the first order in perturbation theory
in the second line.
The above equation reads:
\begin{eqnarray}\label{2pf5}
&&\langle\phi^i(z_1,\bar z_1)\,\phi^j(z_2,\bar z_2)\rangle_{\lambda_A}\nonumber\\
&=&\frac{\langle\phi^i(z_1,\bar z_1)\,\phi^j(z_2,\bar z_2)\rangle_0
+\lambda_A\int d^2z\langle\phi^i(z_1,\bar z_1)\,\mathcal{O}_{d,A}(z,\bar z)
\,\phi^j(z_2,\bar z_2)\rangle+O(\lambda_A^2)}
{1+\lambda_A\int d^2z\langle\mathcal{O}_{d,A}(z,\bar z)\rangle+
O(\lambda_A^2)}\nonumber\\
&=&\langle\phi^i(z_1,\bar z_1)\,\phi^j(z_2,\bar z_2)\rangle_0+\lambda_A\int d^2z
\langle\phi^i(z_1,\bar z_1)\,\mathcal{O}_{d,A}(z,\bar z)\,
\phi^j(z_2,\bar z_2)\rangle+O(\lambda_A^2),
\end{eqnarray}
where $\langle\mathcal{O}_{d,A}(z,\bar z)\rangle=0$.
We therefore obtain:
\begin{equation}\label{2pf6}
\frac{\partial}{\partial\lambda_A}\langle\phi^i(z_1,\bar z_1)\,
\phi^j(z_2,\bar z_2)\rangle_{\lambda_A}=\int d^2z
\langle\phi^i(z_1,\bar z_1)\,\mathcal{O}_{d,A}(z,\bar z)\,
\phi^j(z_2,\bar z_2)\rangle.
\end{equation}
One then needs to evaluate the above three-point function:
\begin{eqnarray}\label{3pf1}
&&\int d^2z\langle\phi^i(z_1,\bar z_1)\,\mathcal{O}_{d,A}(z,\bar z)\,
\phi^j(z_2,\bar z_2)\rangle=
C_{iAj} \int d^2z  \times \\&&
\times \frac{1}{(z_1-z)^{h_i+1-h_j}\,(z-z_2)^{h_j+1-h_i}\,(z_{12})^{h_i+h_j-1}\,
(\bar z_1-\bar z)^{\tilde h_i+1-\tilde h_j}\,(\bar z-\bar z_2)^{\tilde h_j+1-\tilde h_i}\,
(\bar z_{12})^{\tilde h_i+\tilde h_j-1}},\nonumber
\end{eqnarray}
where $C_{iAj}$ are the structure constants, the index ``A" corresponds to the deformation operator.
The integral on the right hand side of the above equation needs to be regularized by putting cutoffs at the
insertion points of the quasi-primary fields: $|z-z_1|>\epsilon$ and $|z-z_2|>\epsilon$.  As discussed in \cite{Eberle:2001jq,eberleth}, one can make the SL(2,$\mathbb{C}$)
transformation:
\begin{equation}\label{sl2c}
y(z)=\frac{z_{12}z}{z_{12}-z},
\end{equation}
which brings the above integral to the form
\be
\int_{|y|=\frac{\epsilon}{|z_{12}|^2}}^{|y|=\frac{1}{\epsilon}}dy d\bar{y} \frac{e^{i\pi (h_j-h_i+1-(\ti h_j-\ti h_j +1))}}{y^{h_j-h_i+1}\bar{y}^{\ti h_j-\ti h_i+1}}
\ee
where we have chosen any branch cuts coming from possible fractional powers of $z_{ij}$ so as to align with the branch cuts coming from possible fractional powers of $\bar{z}_{ij}$, and cancel.  We now make the following comment.  The above integral is a two dimensional integral, and the domain of integration is rotationally symmetric.  This means that the angular part of the integral is unconstrained. This immediately requires that
\bea
h_j-h_i+1 &=&\ti h_j-\ti h_i+1 \nn \\
h_j-\ti h_j& =& h_i-\ti h_i. \label{spinconserve}
\eea
i.e. that the spin of the fields $\phi_i$ and $\phi_j$ should match.  We have not broken the rotation group with the addition of weight $(1,1)$ operators to the action (they are spinless), and so we should expect this to be a respected quantum number in the perturbed theory.

Now we see that there are two different possibilities: 1) $h_i=h_j$, $\ti h_i=\ti h_j$ (hence, $i=j$, or explicitly $\phi^i=\phi^j$ given our
assumption), and 2) $h_i\ne h_j$, $\ti h_i\neq \ti h_j$ ($i\neq j$) \cite{Dijkgraaf:1987jt}.
For $h_i=h_j$ the integral reads:
\begin{eqnarray}\label{3pf2}
&&\int d^2z\langle\phi^i(z_1,\bar z_1)\,\mathcal{O}_{d,A}(z,\bar z)\,
\phi^{i\dagger}(z_2,\bar z_2)\rangle=\nonumber\\
&&\int d^2z\frac{C_{iAi}}{(z_1-z)^{1}\,(z-z_2)^{1}\,(z_{12})^{2h_i-1}\,
(\bar z_1-\bar z)^{1}\,(\bar z-\bar z_2)^{1}\,
(\bar z_{12})^{2\tilde h_i-1}}=\nonumber\\
&&\frac{1}{z_{12}^{2h_i}\,\bar z_{12}^{2\tilde h_i}}\,C_{iAi}\int d^2z
\frac{|z_{12}|^2}{|z_{1}-z|^2\,|z-z_{2}|^2}.
\end{eqnarray}
The SL(2,$\mathbb{C}$) transformation (\ref{sl2c}) simplifies the integral in the third line of the above equation
\begin{eqnarray}\label{3pf3}
&&\frac{C_{iAi}}{z_{12}^{2h_i}\,\bar z_{12}^{2\tilde h_i}}
\int_{|z-z_1|>\epsilon,|z-z_2|>\epsilon} d^2z
\frac{|z_{12}|^2}{|z_{1}-z|^2\,|z-z_{2}|^2}
=\frac{C_{iAi}}{z_{12}^{2h_i}\,\bar z_{12}^{2\tilde h_i}}
\int_{\epsilon<|y|<|z_{12}^2/\epsilon}\frac{d^2y}{|y|^2}\\
&&=\Big(2\pi C_{iAi}\ln(z_{12})+2\pi C_{iAi}\ln(\bar z_{12})
-2\pi C_{iAi}\ln(\epsilon^2)\Big)
\langle\phi^i(z_1,\bar z_1)\,\phi^{i\dagger}(z_2,\bar z_2)\rangle_0.\nonumber
\end{eqnarray}
The $\epsilon$-dependent term which diverges in the limit $\epsilon\to0$ has to be absorbed in the renormalization of $\phi^i$.

For $h_i\ne h_j$ the integral (\ref{3pf1}) reads:
\begin{eqnarray}\label{3pf4}
&&\int d^2z\langle\phi^i(z_1,\bar z_1)\,\mathcal{O}_{d,A}(z,\bar z)\,
\phi^j(z_2,\bar z_2)\rangle= C_{iAj} \int d^2z \times \\&&
\times \frac{1}{(z_1-z)^{h_i+1-h_j}\,(z-z_2)^{h_j+1-h_i}\,(z_{12})^{h_i+h_j-1}\,
(\bar z_1-\bar z)^{\tilde h_i+1-\tilde h_j}\,(\bar z-\bar z_2)^{\tilde h_j+1-\tilde h_i}\,
(\bar z_{12})^{\tilde h_i+\tilde h_j-1}}. \nonumber
\end{eqnarray}
Performing the same SL(2,$\mathbb{C}$) transformation (\ref{sl2c}) we obtain:
\begin{eqnarray}\label{3pf5}
\int d^2z\langle\phi^i(z_1,\bar z_1)\,\mathcal{O}_{d,A}(z,\bar z)\,
\phi^j(z_2,\bar z_2)\rangle
=\frac{|z_{12}|^{2(h_i-h_j)}}{z_{12}^{h_i+h_j}\,\bar z_{12}^{\tilde h_i+\tilde h_j}}\,
\delta_{s_i,s_j}\,C_{iAj}
\int_{\epsilon<|x|<|z_{12}^2/\epsilon}\frac{d^2y}{|y|^{2(h_i-h_j+1)}}\nonumber\\
=-2\pi\,\delta_{s_i,s_j}\,C_{iAj}
\left(\frac{\epsilon^{d_j-d_i}}{d_j-d_i}\frac{1}{z_{12}^{2h_j}\bar z_{12}^{2\tilde h_j}}+
\frac{\epsilon^{d_i-d_j}}{d_i-d_j}\frac{1}{z_{12}^{2h_i}\bar z_{12}^{2\tilde h_i}} \right),
\qquad\qquad\qquad \qquad\qquad\qquad\quad
\end{eqnarray}
where where $s_i=h_i-\tilde h_i$ is the spin and $d_i\equiv h_i+\tilde h_i$ is the
scaling dimension of  $\phi^i$.
In the limit $\epsilon\to0$ the two terms in the last line of the above equation
either diverge or vanish.
Again, the cut-off dependent term is absorbed into the renormalization of $\phi^i$.
Using (\ref{3pf3}) and (\ref{3pf5}) we can now find the appropriate wave function
renormalization to the first order:
\begin{equation}\label{renorm}
\phi^i\to
\phi^i+\lambda\,\pi\,\ln(\epsilon^2)\,C_{iAi}\,\phi^i+\lambda\,2\pi\,
\sum_j\delta_{s_i,s_j}\,\frac{\epsilon^{d_j-d_i}}{d_j-d_i}\,C_{iAj}\,\phi^j.
\end{equation}
After subtracting the infinite parts, we shall compare the finite result (\ref{3pf3})
with what we obtained earlier in (\ref{2der1i}) and find the anomalous dimension to first order:
\begin{equation}\label{2pf7}
\frac{\partial h_i}{\partial\lambda}=-\pi C_{iAi},\quad
\frac{\partial \tilde h_i}{\partial\lambda}=-\pi C_{iAi} \,.
\end{equation}
This is Kadanoff's deformation theory to the first order.
The structure constants $C_{iAi}$, which are the coefficients of the logarithmic
terms $\ln z_{12}$ and $\ln \bar z_{12}$, determine the anomalous dimension of $\phi^i$ to the first order in perturbation theory.
For discussion of deforming the D1-D5 CFT away from the orbifold point at second order in perturbation theory, see \cite{Pakman:2009mi}.

So far we assumed that there is no degeneracy in the conformal weights of quasi primaries $\phi^i$.  To relax this condition, we simply note that the form of the integrals remains unchanged for the cases $h_i=h_j$, $\ti h_i= \ti h_j$ or $h_i\neq h_j$, $\ti h_i\neq \ti h_j$.  Therefore, this does not affect the sector where $h_i \neq h_j$ because this requirement means that the fields are distinct: this part of the computation remains intact.  The only modification is for the $h_i=h_j$ part of the computation, where the coefficient is now $C_{iAj}$.  To find the correction to the anomalous dimensions, we would need to diagonalize $C_{iAj}$ in the entire block of fields with the same conformal dimension.  Rather than considering all operators with the same conformal dimension, one may consider other preserved symmetries, like R-symmetry, to restrict the search.  We could also imagine trying to find the operators iteratively, by taking a given operator $\phi_1$ and finding the operators that this mixes with that have the same conformal weight $\phi_2\cdots \phi_n$.  One would then find all the operators that these operators mix with that have the same conformal dimension, and so on until one finds the full set of fields that mix.  We will further outline how one may attempt to do this in our discussion, in section \ref{discsect}.

\subsection{Four-point functions and factorization channels}\label{meth}
Having constructed the deformation operator, we can start the computation of the
anomalous dimensions of some candidate states of the D1-D5 orbifold CFT.
The states that we consider belong to the non-twist sector of the theory.

Super chiral primary states of the orbifold CFT and their descendants under the
anomaly-free subalgebra of the superconformal algebra make the short multiplets
of the theory \cite{David:2002wn}.
The perturbative orbifold CFT and the supergravity theory are appropriate descriptions
in different parts of the moduli space of the D1-D5 system.
Therefore, if we want to compare the states of the two theories, we have to consider
those which are protected  against corrections as one moves in the moduli space.
The states belonging to the short multiplets of the orbifold CFT do not acquire corrections
as one moves across the moduli space.
These states identify the supergravity modes in the near-horizon geometry of the
D1-D5 system.
The energy and two- and three-point functions of the members of the short multiplets are
not renormalized \cite{Baggio:2012rr}.

We consider one of the protected states of the orbifold CFT in the next section.
This state corresponds to the dilaton in the dual supergravity description.
We evaluate the anomalous dimension of the state to the first order in conformal
perturbation theory and show that it vanishes, as expected.
Then, in the following section, we will consider a non-protected state of the orbifold CFT.
We will investigate the corrections to the conformal weight of our candidate state
as we deform the CFT away from the orbifold point.
We first describe our method of calculation in the remaining of this section.

As discussed earlier in section \ref{cpft}, Kadanoff's deformation theory gives the
anomalous dimensions acquired by the states of a CFT under a small perturbation.
To the first order in conformal perturbation theory, the anomalous dimension of a state
$|\phi^i\rangle$ with non-degenerate conformal weight $(h_i,\bar h_i)$
is given by (\ref{2pf7}):
\begin{equation}
\frac{\partial h_i}{\partial\lambda}=-\pi C_{iAi},\quad
\frac{\partial \tilde h_i}{\partial\lambda}=-\pi C_{iAi},\nonumber
\end{equation}
where $C_{iAi}$ are the structure constants corresponding to the three-point functions
$\langle\phi^i\,\mathcal{O}_{d}\,\phi^{i\dagger}\rangle$.
For the non-twist sector operators that we consider, this correlator vanishes because of a group selection rule.
For CFT states with degenerate conformal weights (such as our candidate states),
one needs to worry about potential first-order operator mixing between $\phi^i$ and other
quasi-primary operators $\phi^j$ with the \emph{same} conformal weight $(h_i,\ti h_i)$.
If there exists such mixing, one has to identify all the operators which $\phi^i$ mixes with,
evaluate $C_{iAj}$, diagonalize the matrix of the structure constants, and find the
change to the conformal weight.

In order to investigate first-order mixing between operators of the same
conformal weight we evaluate the four-point function involving the operator under
consideration, its Hermitian conjugate, and two insertions of the deformation operator:
\begin{equation}\label{iAAi}
\langle\phi^i(z_1,\bar z_1)\,\mathcal{O}_{d}(z_2,\bar z_2)
\,\mathcal{O}_{d}(z_3,\bar z_3)\,\phi^{i\dagger}(z_4,\bar z_4)\rangle.
\end{equation}
We take the coincidence limit as $z_1\to z_2$, $z_3\to z_4$ (or equivalently $z_1\to z_3$, $z_2\to z_4$)
and find the leading singular term and its coefficient.
This singular limit signals intermediate quasi-primary operator(s) which mix with
$\phi^i$ at the first order.
We evaluate the conformal weight of the quasi-primary operator(s) and subtract their
conformal families from the leading order singular limit of the four-point function
(there may be more than one quasi-primary operators $\phi^k$, with the same
conformal dimension, which contribute to this leading singularity.
The sum over $C_{iA}^k$ of  these operators must give the coefficient of the leading
singular term).
Each conformal family is composed of an ancestor quasi-primary operator and
all its descendants under the Virasoro algebra.
After subtracting these intermediate conformal families, we find the remaining leading
order singularity, evaluate the conformal weight of the quasi-primaries of this singular limit,
and subtract their intermediate conformal families.
We continue this procedure until we exhaust all the intermediate conformal families
which mix with $\phi^i$ at the first order in perturbation theory.

We are interested in conformal families whose ancestors $\phi^j$ have the same
conformal weight as $\phi^i$ \footnote{
There are two different types of operators which can have the same conformal weight as
$\phi^i$: quasi-primary operators which are the ancestors of conformal families, and
secondary operators which are the descendants of quasi-primaries under the Virasoro
algebra.
As shown in  Section \ref{dfmn},  it turns out that one needs consider only contributions from quasi-primary (\emph{i.e.,} non-derivative) operators.}. 
These operators contribute to the anomalous dimension of $\phi^i$.
Quasi-primaries which do not have the same conformal weight as $\phi^i$
contribute only to the wave function renormalization at the first order in perturbation theory
(\ref{renorm}), but do not change the conformal dimension.

Computing the four-point function (\ref{iAAi}) and taking its coincidence limits is a
robust way to compute the mixing coefficient of the set of \emph{all} quasi-primaries which
mix with $\phi^i$ at
the first order.
There is a variety of building blocks in the $\mathcal{N}=(4,4)$ D1-D5 orbifold CFT which
we can use to construct candidate quasi-primaries which participate in operator mixing.

\section{Dilaton warm-up}\label{dilaton}
The AdS$_3$/CFT$_2$ correspondence \cite{Maldacena:1997re} matches the 20 exactly
marginal operators of the $\mathcal{N}=(4,4)$ orbifold CFT with the 20 near-horizon
supergravity moduli \cite{David:2002wn}.
Under the correspondence, the six-dimensional dilaton in supergravity is identified with
the exact marginal operator
$\sum_{\kappa=1}^N\partial x^i_{(\kappa)}\bar\partial x_{i\,(\kappa)}$,
where $x^i$, $i\in\{1,2,3,4\}$ are the four real bosonic fields and, as before, $\kappa$
is the copy index.
We refer to this operator as the dilaton operator.
Since the exactly marginal operators are protected from getting corrections, one expects
that they do not acquire an anomalous dimension as one moves away from the orbifold
point.
We check this as a warm up calculation by evaluating the four point function (\ref{iAAi})
for the dilaton operator.

\subsection{Four-point function}\label{fourpf}
We can write the dilaton operator in terms of the complex bosons $X^{\dot AA}$:
\begin{equation}\label{dil}
\sum_{\kappa=1}^N\partial x^i_{(\kappa)}\bar\partial x_{i\,(\kappa)}=
\sum_{\kappa=1}^N-\epsilon_{AB}\,\epsilon_{\dot A\dot B}
\partial X^{\dot AA}_{(\kappa)}\,\bar\partial X^{\dot BB}_{(\kappa)}.
\end{equation}
The dilaton operator is self-conjugate.
The four-point function which we would like to compute is of the form:
\begin{eqnarray}\label{dAAd1}
\Bigg<\bigg(\sum_{\kappa=1}^N-\epsilon_{AB}\,\epsilon_{\dot A\dot B}
\partial X^{\dot AA}_{(\kappa)}\,\bar\partial X^{\dot BB}_{(\kappa)}\bigg)
(z_1,\bar z_1)\quad
\lambda\,\mathcal{O}_d
(z_2,\bar z_2)\times\qquad\qquad\nonumber\\
\times\lambda\,\mathcal{O}_d
(z_3,\bar z_3)\quad
\bigg(\sum_{\kappa^\prime=1}^N-\epsilon_{A^{\prime}B^{\prime}}\,
\epsilon_{\dot A^{\prime}\dot B^{\prime}}
\partial X^{\dot A^{\prime}A^{\prime}}_{(\kappa^{\prime})}\,
\bar\partial X^{\dot B^{\prime}B^{\prime}}_{(\kappa^{\prime})}\bigg)
(z_4,\bar z_4)\Bigg>.
\end{eqnarray}
We use the translational invariance of the correlation function and shift the position
of the deformations operator at $z_3$ to zero.
Defining the new positions as $a_1\equiv z_1-z_3$, $b\equiv z_2-z_3$, and
$a_2\equiv z_4-z_3$, we obtain:
\begin{eqnarray}\label{dAAd2}
\Bigg<\bigg(\sum_{\kappa=1}^N-\epsilon_{AB}\,\epsilon_{\dot A\dot B}
\partial X^{\dot AA}_{(\kappa)}\,\bar\partial X^{\dot BB}_{(\kappa)}\bigg)(a_1,\bar a_1)\quad
\lambda\,\mathcal{O}_d
(b,\bar b)\times
\qquad\qquad\nonumber\\
\times\lambda\,\mathcal{O}_d
(0,0)\quad
\bigg(\sum_{\kappa^\prime=1}^N-\epsilon_{A^{\prime}B^{\prime}}\,
\epsilon_{\dot A^{\prime}\dot B^{\prime}}
\partial X^{\dot A^{\prime}A^{\prime}}_{(\kappa^{\prime})}\,
\bar\partial X^{\dot B^{\prime}B^{\prime}}_{(\kappa^{\prime})}\bigg)
(a_2,\bar a_2)\Bigg>.
\end{eqnarray}
The deformation operator contains a twist-2 operator $\sigma^0_2$ which permutes
two copies of the target space.
Let us denote the two copies as $\kappa=1,2$.
To find the $S_N$ invariant operator we have to sum over the conjugacy classes of 2-cycles.
This brings a combinatorial factor in the definition of the deformation operator
and will be taken into account at the end of this section.
For the moment we consider the $S_N$ non invariant correlator for the
two copies $\kappa=1,2$, compute the four-point
function, and take the coincidence limit to study the operator mixing.
The overall combinatorial coefficient does not affect the arguments here,
but obviously does have to be taken into account when evaluating the corrections to the conformal weight.
To make the notation compact we define
$\phi_{\mathrm{dil}\,(\kappa)}\equiv-\epsilon_{AB}\,\epsilon_{\dot A\dot B}
\partial X^{\dot AA}_{(\kappa)}\,\bar\partial X^{\dot BB}_{(\kappa)}$.
The four-point function (\ref{dAAd2}) is rewritten as:
\begin{eqnarray}\label{dAAd3}
&&\Big<\sum_{\kappa=1}^N\phi_{\mathrm{dil}\,(\kappa)}(a_1,\bar a_1)\;\;
\lambda\mathcal{O}_d
(b,\bar b)\;\;
\lambda\mathcal{O}_d(0,0)\;\;
\sum_{\kappa^\prime=1}^N\phi_{\mathrm{dil}\,(\kappa^{\prime})}
(a_2,\bar a_2)\Big>=\nonumber\\
&&\Big<\sum_{\kappa=1}^2\phi_{\mathrm{dil}\,(\kappa)}(a_1,\bar a_1)\;\;
\lambda\mathcal{O}_d
(b,\bar b)\;\;
\lambda\mathcal{O}_d
(0,0)\;\;
\sum_{\kappa^\prime=1}^2\phi_{\mathrm{dil}\,(\kappa^{\prime})}
(a_2,\bar a_2)\Big>+\nonumber\\
&&\Big<\sum_{\kappa=1}^2\phi_{\mathrm{dil}\,(\kappa)}(a_1,\bar a_1)\;\;
\lambda\mathcal{O}_d
(b,\bar b)\;\;
\lambda\mathcal{O}_d
(0,0)\;\;
\sum_{\kappa^\prime=3}^N\phi_{\mathrm{dil}\,(\kappa^{\prime})}
(a_2,\bar a_2)\Big>+\nonumber\\
&&\Big<\sum_{\kappa=3}^N\phi_{\mathrm{dil}\,(\kappa)}(a_1,\bar a_1)\;\;
\lambda\mathcal{O}_d
(b,\bar b)\;\;
\lambda\mathcal{O}_d
(0,0)\;\;
\sum_{\kappa^\prime=1}^2\phi_{\mathrm{dil}\,(\kappa^{\prime})}
(a_2,\bar a_2)\Big>+\nonumber\\
&&\Big<\sum_{\kappa=3}^N\phi_{\mathrm{dil}\,(\kappa)}(a_1,\bar a_1)\;\;
\lambda\mathcal{O}_d
(b,\bar b)\;\;
\lambda\mathcal{O}_d
(0,0)\;\;
\sum_{\kappa^\prime=3}^N\phi_{\mathrm{dil}\,(\kappa^{\prime})}
(a_2,\bar a_2)\Big>.\quad
\end{eqnarray}
The correlation functions on the third and fourth line of the above equation vanish because
the deformation operators permute copies 1 and 2 of the target space.
The correlation function in the last line factorizes and we obtain:
\begin{eqnarray}\label{dAAd4}
&&\Big<\sum_{\kappa=1}^N\phi_{\mathrm{dil}\,(\kappa)}(a_1,\bar a_1)\;\;
\lambda\mathcal{O}_d
(b,\bar b)\;\;
\lambda\mathcal{O}_d
(0,0)\;\;
\sum_{\kappa^\prime=1}^N\phi_{\mathrm{dil}\,(\kappa^{\prime})}
(a_2,\bar a_2)\Big>=\nonumber\\
&&\Big<\sum_{\kappa=1}^2\phi_{\mathrm{dil}\,(\kappa)}(a_1,\bar a_1)\;\;
\lambda\mathcal{O}_d
(b,\bar b)\;\;
\lambda\mathcal{O}_d+
(0,0)\;\;
\sum_{\kappa^\prime=1}^2\phi_{\mathrm{dil}\,(\kappa^{\prime})}
(a_2,\bar a_2)\Big>+\nonumber\\
&&\Big<\sum_{\kappa=3}^N\phi_{\mathrm{dil}\,(\kappa)}(a_1,\bar a_1)\;\;
\sum_{\kappa^\prime=3}^N\phi_{\mathrm{dil}\,(\kappa^{\prime})}
(a_2,\bar a_2)\Big>
\Big<
\lambda\mathcal{O}_d
(b,\bar b)\;\;
\lambda\mathcal{O}_d
(0,0)
\Big>.
\end{eqnarray}
The factorized correlation function in the last line provides no information about
the mixing between the dilaton operator and other operators of the CFT because it is not singular in the coincidence limits of concern, e.g. $a_1\rightarrow b$.
Thus this term is of no interest for our purpose.
The four-point function in the second line is the one which contains the information
we are after.

\subsection{Mapping from the base to the cover}\label{themap}
Lunin-Mathur (LM) technology \cite{Lunin:2000yv,Lunin:2001pw} for symmetric orbifolds allows computation of correlation functions involving twist sector operators by lifting the correlator to the covering surface. The insight resides in choosing appropriate maps which correctly lift the ramified points -- the places where twist operators are inserted -- to the covering surface. This ensures the right number of images in the cover both for twist and non-twist sector operators. In particular, LM showed how to normalize the bare twist contribution to the correlation function properly by evaluating the Liouville factor corresponding to the conformal map. This procedure is necessary to get the boundary conditions right for the ramified points.  LM technology also makes clear how to regularize integrals, in essence by cutting out disks.

In our companion paper \cite{companion}, we further developed LM technology for symmetric orbifolds to the non-twist sector. This turned out to be a very natural extension of their methods. We needed the generalization in order to be able to calculate correlation functions involving both twist and non-twist  operators. To illustrate the techniques, we worked through two examples: (i) excitations of twist operators by modes of fields unaffected by twist operators, and (ii) non-twist  operators. We refer readers interested in details to \cite{companion}.

For the task at hand, we want to evaluate our four-point correlator obtained in the last subsection by using generalized LM technology. The correlation function has two insertion of twist-2 operators at $z=b$ and $z=0$.
We will first write the map from the base space to the covering sphere and then use the map to pass to the cover and compute the correlator.

We wrote down the map from the base space to the covering surface for two insertions of twist-$n$ operators in \cite{companion}. Here we have $n=2$ and the map is of the form
\begin{equation}\label{map2-1}
z=b\,\frac{t^2}{2t-1}.
\end{equation}
In the vicinity of the two insertion points we have:
\begin{eqnarray}\label{map2-2}
z\to b,\;t\to1,&& (z-b)\approx b_1(t-1)^2+b_1^\prime(t-1)^3+O((t-1)^4),\nonumber\\
z\to0,\;t\to0,&& z\approx b_0\,t^2+b_0^\prime\,t^3+O(t^4),
\end{eqnarray}
where $b_1=b$ and $b_0=-b$.
A generic point in the base space $a_k$ has two images on the covering surface which
we refer to them as $t_{\pm k}$ and are given by:
\begin{equation}
t_{\pm k}=\frac{1}{b}\left(a_k\pm\sqrt{a_k(a_k-b)}\right).
\end{equation}
In the vicinity of a generic point the map is of the form:
\begin{equation}\label{map2-3}
(z-a_k)\approx\xi_{\pm k}\,(t-t_{\pm k})+\xi_{\pm k}^\prime\,(t-t_{\pm k})^2+O((t-t_{\pm k})^3),
\end{equation}
where
\begin{equation}\label{xi}
\xi_{\pm k}=\left(\frac{dz}{dt}\right)\bigg|_{t=t_{\pm k}}=
\frac{2\,a_k\,(a_k-b)}{b\,t_{\pm k}\,(t_{\pm k}-1)}.
\end{equation}

There are two different contributions to the four-point function (\ref{dAAd4}): a bare-twist part and a mode-insertion part.
The bare twist part contains the two insertions of $\sigma^0_2$ twists, and is computed using the Liouville action.
Using the Lunin-Mathur method, the normalized two-point function is
\begin{equation}\label{2pf2}
\frac{\langle\sigma_2^0(b,\bar b)\,\sigma_2^0(0,0)\rangle}
{\langle\sigma_2^0(1,1)\,\sigma_2^0(0,0)\rangle}=|b|^{-4\,h_{\sigma^0_2}},
\end{equation}
where $h_{\sigma^0_2}=3/8$.

We next evaluate the mode insertion part of the correlator.
In section \ref{dfmn} we evaluated the complete deformation operator on the cover
(\ref{defool}).
We now lift the dilaton operator explicitly to the covering surface.
We have:
\begin{eqnarray}\label{dilco1}
\alpha^{\dot AA}_{-1\,{(\kappa)}}\,\tilde\alpha^{\dot BB}_{-1\,{(\kappa)}}=
\oint_{a_k}\frac{dz}{2\pi i}\oint_{\bar a_k}\frac{d\bar z}{2\pi i}\,
(z-a_k)^{1-1-1}\;(\bar z-\bar a_k)^{1-1-1}\,
\partial X^{\dot AA}_{(\kappa)}(z)\,\bar\partial X^{\dot BB}_{(\kappa)}(\bar z)
\qquad
\nonumber\\
\to\oint_{t_{\pm k}}\frac{dt}{2\pi i}\left(\frac{dz}{dt}\right)^{1-1}(z-a_k)^{-1}
\oint_{\bar t_{\pm k}}\frac{d\bar t}{2\pi i}\left(\frac{d\bar z}{d\bar t}\right)^{1-1}
(\bar z-\bar a_k)^{-1}
\partial X^{\dot AA}(t)\,\bar\partial X^{\dot BB}(\bar t)
\qquad\;\;
\nonumber\\
=\oint_{t_{\pm k}}\frac{dt}{2\pi i}\,
\frac{\partial X^{\dot AA}(t)}{\bigg[\xi_{\pm k}(t-t_{\pm k})\Big(1+O(t-t_{\pm k})\Big)\bigg]}
\oint_{\bar t_{\pm k}}\frac{d\bar t}{2\pi i}\,
\frac{\bar\partial X^{\dot BB}(\bar t)}{\bigg[\bar\xi_{\pm k}(\bar t-\bar t_{\pm k})
\Big(1+O(\bar t-\bar t_{\pm k})\Big)\bigg]}
\nonumber\\
=\frac{1}{|\xi_{\pm k}|^{2}}\partial X^{\dot AA}(t_{\pm k})\,
\bar\partial X^{\dot BB}(\bar t_{\pm k}),\label{dxdbxsimple}
\qquad\qquad\qquad\qquad\qquad\qquad\qquad\qquad\qquad\qquad\qquad\quad
\end{eqnarray}
where the arrow in the second line implies that we passed from the base to the
cover using the local map (\ref{map2-3}).  The above computation has a relatively simple form because the operator $\pa X \pab X$ does not need to be regulated, and transforms like a tensor under conformal transformations.
The dilaton operator is then given by:
\begin{equation}\label{dilco2}
\phi^t_{\mathrm{dil}}(t_{\pm k},\bar t_{\pm k})\equiv
\frac{-1}{|\xi_{\pm k}|^{2}}\,\epsilon_{\dot A\dot B}\,\epsilon_{AB}\,
\partial X^{\dot AA}(t_{\pm k})\,\bar\partial X^{\dot BB}(\bar t_{\pm k}),
\end{equation}
where the superscript $t$ denotes the fact that this is an operator in the covering surface.
Following \cite{companion}, 
the mode insertion part of the
four-point function is obtained by summing over the images of the non-twist insertions:
\begin{equation}\label{dAAdmo1}
\sum_{t_{\pm1}}\sum_{t_{\pm2}}
\Big<\phi^t_{\mathrm{dil}}(t_{\pm 1},\bar t_{\pm 1})\;\;
\lambda\mathcal{O}^t_d
(1,1)\;\;
\lambda\mathcal{O}^t_d
(0,0)\;\;
\phi^t_{\mathrm{dil}}(t_{\pm 2},\bar t_{\pm 2})\Big>,
\end{equation}
where $t_{\pm 1}$ and $t_{\pm 2}$ are the images of $a_1$ and $a_2$, respectively,
under the map (\ref{map2-1}).
We first evaluate the four-point function in the above equation and then sum over the
images.

Using the dilaton operator (\ref{dilco2}), the deformation (\ref{defool}), the map
coefficients (\ref{xi}), and taking into account the bare twist contribution (\ref{2pf2}),
we find:

\begin{eqnarray}\label{dAAdmo2}
&&|b|^{-\frac32}\,\Big<\phi^t_{\mathrm{dil}}(t_{\pm 1},\bar t_{\pm 1})\;
\lambda\mathcal{O}^t_d
(1,1)\;
\lambda\mathcal{O}^t_d
(0,0)\;
\phi^t_{\mathrm{dil}}(t_{\pm 2},\bar t_{\pm 2})\Big>=
\nonumber\\
&&\frac{\lambda^2}{4}\,|a_1|^{-2}\,|a_2|^{-2}\,|a_1-b|^{-2}\,|a_2-b|^{-2}\Bigg\{
8\,\frac
{|t_{\pm1}|^2\,|t_{\pm1}-1|^2\,|t_{\pm2}|^2|\,t_{\pm2}-1|^2}
{|t_{\pm1}-t_{\pm2}|^4}
\nonumber\\
&&+2\frac
{(t_{\pm1})\,(t_{\pm2}-1)\,|t_{\pm1}-1|^2\,|t_{\pm2}|^2}
{(\bar t_{\pm1})\,(\bar t_{\pm2}-1)\,(t_{\pm1}-t_{\pm2})^2}
+2\frac
{(t_{\pm1}-1)\,(t_{\pm2})\,|t_{\pm1}|^2\,|t_{\pm2}-1|^2}
{(\bar t_{\pm1}-1)\,(\bar t_{\pm2})\,(t_{\pm1}-t_{\pm2})^2}
\qquad\qquad
\nonumber\\
&&+2\frac
{(\bar t_{\pm1})\,(\bar t_{\pm2}-1)\,|t_{\pm1}-1|^2\,|t_{\pm2}|^2}
{(t_{\pm1})\,(t_{\pm2}-1)\,(\bar t_{\pm1}-\bar t_{\pm2})^2}
+2\frac
{(\bar t_{\pm1}-1)\,(\bar t_{\pm2})\,|t_{\pm1}|^2\,|t_{\pm2}-1|^2}
{(t_{\pm1}-1)\,(t_{\pm2})\,(\bar t_{\pm1}-\bar t_{\pm2})^2}
\qquad\quad\;\,
\nonumber\\
&&+
2\frac
{|t_{\pm1}|^2\,|t_{\pm2}-1|^2}
{|t_{\pm1}-1|^2\,|t_{\pm2}|^2}
+2\frac
{|t_{\pm1}-1|^2\,|t_{\pm2}|^2}
{|t_{\pm1}|^2\,|t_{\pm2}-1|^2}
\nonumber\\
&&-
\frac
{(t_{\pm1})\,(t_{\pm2}-1)\,(\bar t_{\pm1}-1)\,(\bar t_{\pm2})}
{(t_{\pm1}-1)\,(t_{\pm2})\,(\bar t_{\pm1})\,(\bar t_{\pm2}-1)}
-\frac
{(t_{\pm1}-1)\,(t_{\pm2})\,(\bar t_{\pm1})\,(\bar t_{\pm2}-1)}
{(t_{\pm1})\,(t_{\pm2}-1)\,(\bar t_{\pm1}-1)\,(\bar t_{\pm2})}
\Bigg\}.
\end{eqnarray}
We normalize the four point function by two-point functions as in \cite{companion}: 
\begin{equation}\label{dAAdmo3}
\frac{\Big<\phi^t_{\mathrm{dil}}(t_{\pm 1},\bar t_{\pm 1})\;
\lambda\mathcal{O}^t_d
(1,1)\;\;
\lambda\mathcal{O}^t_d
(0,0)\;\;
\phi^t_{\mathrm{dil}}(t_{\pm 2},\bar t_{\pm 2})\Big>}
{\big<\phi^t_{\mathrm{dil}}(0,0)\;\phi^t_{\mathrm{dil}}(1,1)\big>\;\;
\Big<
\mathcal{O}^t_d
(0,0)\;
\mathcal{O}^t_d
(1,1)\Big>}.
\end{equation}
(The normalization of the bare twist contribution has been accounted for in (\ref{2pf2})).
The two-point functions are found to be
\begin{equation}\label{defdef}
\Big<
\mathcal{O}^t_d
(0,0)\;\;
\mathcal{O}^t_d
(1,1)\Big>=
8,
\end{equation}
\begin{equation}\label{deldel}
\big<\phi^t_{\mathrm{dil}}(0,0)\;\;\phi^t_{\mathrm{dil}}(1,1)\big>=4.
\end{equation}

In the above, and in the following, note that  there are combinatoric normalization factors coming from summing over conjugacy classes of 2-cycles. This gives the 3-point function an overall factor of $1/\sqrt{N}$ and the 4-point function an overall factor of $1/N$. We have chosen to suppress these combinatoric factors here for notational clarity: it is always clear where to put them back in afterwards. We use the two-point function as a guide for how to normalize; for further details we refer the reader to the second example in our companion work \cite{companion}. After the dust settles, these $N$-dependent combinatoric factors turn out not to alter the mixing coefficients, and so we may safely ignore them until we evaluate anomalous dimensions.   

\subsection{Summing over images}
We sum over the images of the insertion points of the non-twist operators on the covering
surface to compute the complete correlation function
\begin{eqnarray}\label{dAAdmo4}
\sum_{t_{\pm1}}\sum_{t_{\pm2}}
\Big<\phi^t_{\mathrm{dil}}(t_{\pm 1},\bar t_{\pm 1})\;\;
\lambda\mathcal{O}^t_d
(1,1)\;\;
\lambda\mathcal{O}^t_d
(0,0)\;\;
\phi^t_{\mathrm{dil}}(t_{\pm 2},\bar t_{\pm 2})\Big>=\nonumber\\
\bigg<\Big(\phi^t_{\mathrm{dil}}(t_{+1},\bar t_{+1})+
\phi^t_{\mathrm{dil}}(t_{-1},\bar t_{-1})\Big)\;
\lambda\mathcal{O}^t_d
(1,1)\times
\qquad\qquad\qquad\quad
\nonumber\\
\times\lambda\mathcal{O}^t_d
(0,0)\;\;
\Big(\phi^t_{\mathrm{dil}}(t_{+2},\bar t_{+2})+
\phi^t_{\mathrm{dil}}(t_{-2},\bar t_{-2})\Big)\Big>.
\qquad\qquad
\end{eqnarray}
Inserting the four-point function evaluated in the previous section
(\ref{dAAdmo2}), the complete normalized four-point function reads:
\begin{eqnarray}\label{dAAdtot}
&&\Big<\sum_{\kappa=1}^2\phi_{\mathrm{dil}\,(\kappa)}(a_1,\bar a_1)\;\;
\lambda\mathcal{O}_d+
(b,\bar b)\;\;
\lambda\mathcal{O}_d
(0,0)\;\;
\sum_{\kappa^\prime=1}^2\phi_{\mathrm{dil}\,(\kappa^{\prime})}
(a_2,\bar a_2)\Big>=
\nonumber\\
&&\frac{\lambda^2}{2}\,\bigg\{\,\frac{2}{|a_1-a_2|^4\,|b|^4}+\nonumber\\
&&+2^{-1}\,
\frac{(a_1-b)^{\frac12}\,a_2^{\frac12}}{a_1^{\frac12}\,(a_2-b)^{\frac12}}\;
\frac{\left(1+\frac{a_1\,(a_2-b)}{(a_1-b)\,a_2}\right)}{(a_1-a_2)^2\,b^2}\;\;
\frac{(\bar a_1-\bar b)^{\frac12}\,\bar a_2^{\frac12}}
{\bar a_1^{\frac12}\,(\bar a_2-\bar b)^{\frac12}}\;
\frac{\left(1+\frac{\bar a_1\,(\bar a_2-\bar b)}
{(\bar a_1-\bar b)\,\bar a_2}\right)}{(\bar a_1-\bar a_2)^2\,\bar b^2}+
\nonumber\\
&&+2^{-3}\,
\frac{\left(1+\frac{a_1\,(a_2-b)}{(a_1-b)\,a_2}\right)}
{a_1^{\frac32}\,(a_1-b)^{\frac12}\,a_2^{\frac12}\,(a_2-b)^{\frac32}}\;\;
\frac{(\bar a_1-\bar b)^{\frac12}\,\bar a_2^{\frac12}}
{\bar a_1^{\frac12}\,(\bar a_2-\bar b)^{\frac12}}\;
\frac{\left(1+\frac{\bar a_1\,(\bar a_2-\bar b)}
{(\bar a_1-\bar b)\,\bar a_2}\right)}{(\bar a_1-\bar a_2)^2\,\bar b^2}+
\nonumber\\
&&+2^{-3}\,
\frac{(a_1-b)^{\frac12}\,a_2^{\frac12}}{a_1^{\frac12}\,(a_2-b)^{\frac12}}\;
\frac{\left(1+\frac{a_1\,(a_2-b)}{(a_1-b)\,a_2}\right)}{(a_1-a_2)^2\,b^2}\;\;
\frac{\left(1+\frac{\bar a_1\,(\bar a_2-\bar b)}{(\bar a_1-\bar b)\,\bar a_2}\right)}
{\bar a_1^{\frac32}\,(\bar a_1-\bar b)^{\frac12}\,\bar a_2^{\frac12}\,
(\bar a_2-\bar b)^{\frac32}}+
\nonumber\\
&&+2^{-3}\,
a_1^{-\frac12}\,(a_1-b)^{-\frac32}\,a_2^{-\frac32}\,(a_2-b)^{-\frac12}\;\;
\bar a_1^{-\frac12}\,(\bar a_1-\bar b)^{-\frac32}\,
\bar a_2^{-\frac32}\,(\bar a_2-\bar b)^{-\frac12}+
\nonumber\\
&&+2^{-3}\,a_1^{-\frac32}\,(a_1-b)^{-\frac12}\,a_2^{-\frac12}\,(a_2-b)^{-\frac32}\;\;
\bar a_1^{-\frac32}\,(\bar a_1-\bar b)^{-\frac12}\,
\bar a_2^{-\frac12}\,(\bar a_2-\bar b)^{-\frac32}+
\nonumber\\
&&-2^{-4}\,
a_1^{-\frac12}\,(a_1-b)^{-\frac32}\,a_2^{-\frac32}\,(a_2-b)^{-\frac12}\;\;
\bar a_1^{-\frac32}\,(\bar a_1-\bar b)^{-\frac12}\,
\bar a_2^{-\frac12}\,(\bar a_2-\bar b)^{-\frac32}+
\nonumber\\
&&-2^{-4}\,a_1^{-\frac32}\,(a_1-b)^{-\frac12}\,a_2^{-\frac12}\,(a_2-b)^{-\frac32}\;\;
\bar a_1^{-\frac12}\,(\bar a_1-\bar b)^{-\frac32}\,
\bar a_2^{-\frac32}\,(\bar a_2-\bar b)^{-\frac12}\bigg\}.
\end{eqnarray}
One may take this result and express it in terms of cross ratios, showing that it does in fact fit the correct form for a four-point function of quasi-primary fields.

\subsection{Lack of operator mixing}\label{opmxdil}
Having constructed the four-point correlation function, we can now take the coincidence limit
($a_1,\bar a_1)\to(0,0)$ and $(a_2,\bar a_2)\to (b,\bar b)$, and consider the leading singularity.
The OPE of two quasi-primary operators $O_1$ and $O_2$ has the form \cite{DiFrancesco:1997nk}
\begin{equation}\label{ope}
O_1(z,\bar z)\,O_2(0,0)=\sum_p\sum_{\{k,\tilde k\}}
C^{p\,\{k,\tilde k\}}_{12}z^{h_p-h_1-h_2+K}\,
\bar z^{\tilde h_p-\tilde h_1-\tilde h_2+\tilde K}O_p^{\{k,\tilde k\}}(0,0),
\end{equation}
where $O_p$ is a quasi primary, $\{k,\tilde k\}$ denotes a collection of indices $k_i$ and
$\tilde k_i$ which correspond to the descendant states, and $K\equiv\sum_ik_i$,
$\tilde K\equiv\sum_i\tilde k_i$.
The index $p$ accounts for all conformal families which participate in the OPE.
Taking the coincidence limit, the holomorphic part of the four-point function scales as
$a_1^{-3/2}\,(a_2-b)^{-3/2}$.
According to (\ref{ope}), $h_1=h_2=1$ in our case and the quasi-primary $O_p$ has
conformal weight $h=1/2$.
All the descendants of this conformal family have half-integer weights.
Subleading singularities also have half-integer conformal weights.
Singularities corresponding to mixing with $h=\tilde h=1$ quasi-primary operators
are absent in the coincidence limits of the four-point function.
Thus $C_{iAj}=0$, where $j$ corresponds to any weight (1,1) quasi-primary.
As discussed earlier, the structure constant $C_{iAi}$ also vanishes because of a group selection rule since it corresponds to the insertion of only one twist-2 operator in the base.
The fact that $C_{iAi}=0$ and $C_{iAj}=0$ indicates that the dilaton operator does not acquire an anomalous dimension at the first order in perturbation theory, as expected.

\section{Lifting of a string state}\label{lift}
In this section we consider a string state of the superconformal algebra which is not
protected against corrections as one deforms the theory away from the orbifold point.
We study operator mixing at the first order and analyze lifting of the string state.
The non-twist sector has low-lying string states which are lifted under deformation. Here we address the interaction between the twist and the non-twist sector.
The string state that we consider has the general form:
\begin{equation}\label{sgsgen}
\partial X^a_{m\,(\kappa_1)}\,\partial X^b_{n\,(\kappa_2)}\,
\bar\partial X^c_{k\,(\kappa_3)}\,\bar\partial X^d_{l\,(\kappa_4)}\,|0\rangle_{R},
\end{equation}
where $\kappa_i$ corresponds to the copy of the target space and $|0\rangle_R$
is a Ramond-Ramond ground state.
We make a simple choice and set the modes $n=m=k=l=-1$, and choose the excitations to
belong to the same copy of the target space.
Our string state is given by
\begin{equation}\label{xxxx}
\sum_{\kappa=1}^{N}\phi_{\mathrm{st}\,(\kappa)}\,|0\rangle_R\equiv
\delta_{ab}\,\delta_{cd}
\sum_{\kappa=1}^{N}
\partial X^{a}_{-1\,(\kappa)}\,\partial X^{b}_{-1\,(\kappa)}\,
\bar\partial X^{c}_{-1\,(\kappa)}\,\bar\partial X^{d}_{-1\,(\kappa)}\,|0\rangle_{R}.
\end{equation}
The state is a singlet under the internal $SU(2)_1\times SU(2)_2$.

Physical states of the D1-D5 system are in the Ramond sector where the fermions have
periodic boundary conditions around the circle $S^1$.
We use spectral flow transformation \cite{Schwimmer:1986mf}
to relate the correlation functions in the Ramond sector to correlation functions in the NS sector.
The four-point correlation function that we would like to evaluate is
\begin{equation}\label{4pfrsec}
_R\langle0|\sum_{\kappa=1}^{N}\phi_{\mathrm{st}\,(\kappa)}\;\;
\lambda\mathcal{O}_d
(z,\bar z)\;\;
\lambda\mathcal{O}_d
(z^{\prime},\bar z^\prime)\;\;
\sum_{\kappa^\prime=1}^{N}\phi_{\mathrm{st}\,(\kappa^\prime)}|0\rangle_R.
\end{equation}
Let us choose the Ramond-Ramond ground state which has conformal weight
$(h,\bar h)=(1/4,1/4)$ and $R$-charge $(1/2,1/2)$ under $SU(2)_L\times SU(2)_R$.
Under performing a spectral flow transformation
with parameter $\alpha=-1$, this
Ramond ground state is mapped into the NS ground state.
The bosons are not affected under the spectral flow.
The deformation operator is also mapped into itself as we will now show.

Operators of the CFT which could be written as pure exponentials in the context of the bosonized language transform as
\begin{equation}\label{sfexp}
\phi(z)\to z^{-\alpha m}\,\phi(z),
\end{equation}
under the spectral flow with spectral flow parameter $\alpha$
\cite{Avery:2009tu, Avery:2010er}.
Here $m$ is the charge under $SU(2)_L$.
The same transformation holds for anti-holomorphic operators.
Super chiral and anti-chiral primary operators are represented by pure exponentials in the bosonized
language \cite{Lunin:2001pw}.
The deformation operator contains the super chiral primary $\sigma_2^{++}$ with
$R$-charge $(1/2,1/2)$ under $SU(2)_L\times SU(2)_R$.
The holomorphic part of $\sigma_2^{++}$ thus transforms as
$\sigma_2^+(z)\to z^{\alpha/2}\,\sigma_2^+(z)$ under the spectral flow.
The supercurrents transform as $G^{-A}(z)\to z^{-\alpha/2}\,G^{-A}(z)$.
We then have
\begin{eqnarray}\label{sfdef}
G^{-A}_{-\frac12}(z^\prime)\,\sigma_2^+(z)&=&\oint_{z} \frac{dz^\prime}{2\pi i}\,
G^{-A}(z^{\prime})\,\sigma_2^+(z)\to
\oint_{z} \frac{dz^\prime}{2\pi i}\,
z^{\prime-\frac\alpha2}\,G^{-A}(z^{\prime})\,z^{\frac\alpha2}\,\sigma_2^+(z)\nonumber\\
&=&\oint_{z} \frac{dz^\prime}{2\pi i}\,\Big(1-\frac\alpha2(z^\prime-z)z^{-1}+\cdots\Big)\,
G^{-A}(z^{\prime})\,\sigma_2^+(z)\nonumber\\
&=&G^{-A}_{-\frac12}(z^\prime)\,\sigma_2^+(z)
-\frac\alpha2\,z^{-1}\,G^{-A}_{+\frac12}(z^\prime)\,\sigma_2^+(z)+\cdots,
\end{eqnarray}
where $``\cdots"$ denotes higher positive modes of the supercurrent.
Since positive modes of the supercurrent annihilate super chiral primaries, only
the first term in the third line of the above equation is non-vanishing and
the deformation operator is not affected under the spectral flow.

Under the spectral flow with $\alpha=-1$ the physical problem in the Ramond sector is
mapped into a computation in the NS sector.
The four-point correlation function that we evaluate is
\begin{equation}\label{4pfnssec}
_{NS}\langle0|\sum_{\kappa=1}^{N}\phi_{\mathrm{st}\,(\kappa)}\;\;
\lambda\mathcal{O}_d
(z,\bar z)\;\;
\lambda\mathcal{O}_d
(z^{\prime},\bar z^\prime)\;\;
\sum_{\kappa^\prime=1}^{N}\phi_{\mathrm{st}\,(\kappa^\prime)}|0\rangle_{NS}.
\end{equation}
As discussed in subsection (\ref{fourpf}), four-point functions which factorize are of
no interest for our purpose because they have no information about operator mixing.
Let us denote the two copy indices of the target space which are twisted under the deformation
operator as $\kappa=1,2$.
Then the four-point function in which we are interested contains
$\sum_{\kappa=1}^{2}\phi_{\mathrm{st}\,(\kappa^\prime)}$.
We will next explicitly compute $\phi_{\mathrm{st}}$ on the covering surface and then
evaluate the correlation function.

\subsection{Passing to the covering surface}\label{stpass}
The four-point correlation function (\ref{4pfnssec}) has two insertions of twist-2
operators in the base space.
The map from the base to the cover is thus the same as in the previous section
(\ref{map2-1}).
We set the insertion points of the two deformations at $z=0$ and $z=b$ and the
insertion points of $\phi_{\mathrm{st}}$ at $z=a_1$ and $z=a_2$, as before.
The bare-twist contribution to the correlation function is again the normalized two-point
function of twist-2 operators (\ref{2pf2}) and is given by $|b|^{-3/2}$.
To evaluate the mode insertion contribution to the four-point function, we lift the operators
to the covering sphere.

The operator
$\phi_{\mathrm{st}}=\partial X^{a}_{-1\,(\kappa)}\,\partial X_{a,\,-1\,(\kappa)}\,
\bar\partial X^{b}_{-1\,(\kappa)}\,\bar\partial X_{b,\,-1\,(\kappa)}(z,\bar z)$ is a quasi-primary operator with conformal weight $(2,2)$.
Passing to the covering surface, we obtain
\begin{eqnarray}\label{xxco}
&& \partial X^{a}_{-1\,{(\kappa)}}\,\partial X^{b}_{-1\,{(\kappa)}}=
\oint_{a_k}\frac{dz_1}{2\pi i}\,
(z_1-a_k)^{1-1-1}\,\partial X^{a}_{(\kappa)}(z_1)
\oint_{a_k}\frac{dz_2}{2\pi i}\,
(z_2-a_k)^{1-1-1}\,\partial X^{b}_{(\kappa)}(z_2)
\nonumber\\
&& \to\oint_{t_{\pm k}}\frac{dt_1}{2\pi i}\left(\frac{dz_1}{dt_1}\right)^{1-1}(z_1-a_k)^{-1}\,
\partial X^{a}(t_1)
\oint_{t_{\pm k}}\frac{dt_2}{2\pi i}\left(\frac{dz_2}{dt_2}\right)^{1-1}(z_2-a_k)^{-1}\,
\partial X^{b}(t_2)
\nonumber\\
&& =\oint_{t_{\pm k}}\frac{dt_1}{2\pi i}\,
\frac{\partial X^{a}(t_1)}
{
\xi_{\pm k}(t_1-t_{\pm k})\Big(1+O(t_1-t_{\pm k})\Big)
}
\oint_{t_{\pm k}}\frac{dt_2}{2\pi i}\,
\frac{\partial X^{b}(t_2)}{
\xi_{\pm k}(t_2-t_{\pm k})
\Big(1+O(t_2-t_{\pm k})\Big)
}
\nonumber\\
&&=\oint_{t_{\pm k}}\frac{dt_1}{2\pi i}\,
\frac{\partial X^{a}(t_1)}
{
\xi_{\pm k}(t_1-t_{\pm k})\Big(1+O(t_1-t_{\pm k})\Big)
}
\,\frac{1}{\xi_{\pm k}}\partial X^{b}(t_{\pm k})
\nonumber\\
&&=\oint_{t_{\pm k}}\frac{dt_1}{2\pi i}\,\frac{1}{\xi_{\pm k}^2}
\frac{\Big[1+\eta_1(t_1-t_{\pm k})+\eta_2(t_1-t_{\pm k})^2+O\Big((t_1-t_{\pm k})^3\Big)\Big]}
{t_1-t_{\pm k}}\,\partial X^{a}(t_1)\,\partial X^{b}(t_{\pm k})
\nonumber\\
&&=\frac{1}{\xi_{\pm k}^2}\left(\partial X^{a}\,\partial X^{b}(t_{\pm k})
-\,\frac{\delta^{ab}}{4(t_{\pm k})^2(t_{\pm k}-1)^2}\right), \label{stringstatelift}
\end{eqnarray}
where the arrow in the second line denotes that we passed from the base to the
cover using the map which locally looks like (\ref{map2-3}), and $\eta_1$ and $\eta_2$
are coefficients obtained from expanding the map.  The form of this operator may again be expected because $:{\pa}X{\pa}X:$ needs to be regulated, and this regulation causes this object to not transform as a tensor.  However, the transformation properties of this object for finite transformations can be easily determined, and one recognizes the the Schwarzian derivative
\be
\left\{f(t),t\right\}=-\frac{3}{2} \frac{1}{(t-1)^2t^2}
\ee
of the map (\ref{map2-1}) appearing in (\ref{stringstatelift}) .
The same relation is obtained for the anti-holomorphic part of the operator.
The operator has the two-point function
$\langle\phi_{\mathrm{st}}(0,0)\,\phi_{\mathrm{st}}(1,1)\rangle=2^8$, which is
used to normalize the four-point function.
Using the above equation (\ref{xxco}), the deformation (\ref{defool}), and taking into
account the bare twist contribution, the four-point function normalized by the two-point
functions is evaluated:
\begin{eqnarray}\label{stAAst1}
&&|b|^{-\frac32}\,\Big<\phi^t_{\mathrm{st}}(t_{\pm 1},\bar t_{\pm 1})\;\;
\lambda\mathcal{O}^t_d
(1,1)\;\;
\lambda\mathcal{O}^t_d
(0,0)\;
\phi^t_{\mathrm{st}}(t_{\pm 2},\bar t_{\pm 2})\Big>=
\\
&&\frac{|b|^4}{2^{16}}\,\lambda^2\,
|a_1|^{-4}\,|a_2|^{-4}\,|a_1-b|^{-4}\,|a_2-b|^{-4}\times
\nonumber\\
&&\times\Bigg\{5
+4\,\frac{(t_{\pm1}-1)^2(t_{\pm2})^2}{(t_{\pm1}-t_{\pm2})^2}
+4\,\frac{(t_{\pm1})^2(t_{\pm2}-1)^2}{(t_{\pm1}-t_{\pm2})^2}
+8\,\frac{(t_{\pm1})^2(t_{\pm1}-1)^2(t_{\pm2})^2(t_{\pm2}-1)^2}
{(t_{\pm1}-t_{\pm2})^4}
\Bigg\}\times\nonumber\\
&&\times\Bigg\{5
+4\,\frac{(\bar t_{\pm1}-1)^2(\bar t_{\pm2})^2}{(\bar t_{\pm1}-\bar t_{\pm2})^2}
+4\,\frac{(\bar t_{\pm1})^2(\bar t_{\pm2}-1)^2}{(\bar t_{\pm1}-\bar t_{\pm2})^2}
+8\,\frac{(\bar t_{\pm1})^2(\bar t_{\pm1}-1)^2(\bar t_{\pm2})^2(\bar t_{\pm2}-1)^2}
{(\bar t_{\pm1}-\bar t_{\pm2})^4}
\Bigg\}.\nonumber
\end{eqnarray}

\subsection{Image sums}\label{stsumm}
We finally sum over the images of the insertion points of the two non-twist operators
and compute the complete normalized four-point function.
To make the notation compact, we write down the result in terms of the cross ratio,
$R={a_1(a_2-b)}/{(a_2(a_1-b))}$.
We obtain:
\begin{eqnarray}\label{stAAsttot}
&&\Big<\sum_{\kappa=1}^2\phi_{\mathrm{st}\,(\kappa)}(a_1,\bar a_1)\;\;
\lambda\mathcal{O}_d
(b,\bar b)\;\;
\lambda\mathcal{O}_d
(0,0)\;\;
\sum_{\kappa^\prime=1}^2\phi_{\mathrm{st}\,(\kappa^{\prime})}
(a_2,\bar a_2)\Big>=\nonumber\\
&&\frac{|b|^4}{2^{12}}\,\lambda^2\,|a_1|^{-4}\,|a_2|^{-4}\,|a_1-b|^{-4}\,|a_2-b|^{-4}\Bigg\{\frac{25}{4}+
5\,\bigg(\frac{(R+1)^2}{(R-1)^2}+\frac{(\bar R+1)^2}{(\bar R-1)^2}\bigg)\nonumber\\
&&+16\,\frac{R^{\frac12}(R+1)}{(R-1)^2}\;\frac{\bar R^{\frac12}(\bar R+1)}{(\bar R-1)^2}+
4\,\frac{(R+1)^2}{(R-1)^2}\;\frac{(\bar R+1)^2}{(\bar R-1)^2}\nonumber\\
&&+10\,\bigg(\frac{R\,(R^2+6R+1)}{(R-1)^4}+
\frac{\bar R\,(\bar R^2+6\bar R+1)}{(\bar R-1)^4}\bigg)\nonumber\\
&&+64\,\bigg(
\frac{R^{\frac12}(R+1)}{(R-1)^2}\;\frac{\bar R^{\frac32}(\bar R+1)}{(\bar R-1)^4}+
\frac{R^{\frac32}(R+1)}{(R-1)^4}\;\frac{\bar R^{\frac12}(\bar R+1)}{(\bar R-1)^2}\bigg)
\nonumber\\
&&+8\,\bigg(\frac{(R+1)^2}{(R-1)^2}\;\frac{\bar R\,(\bar R^2+6\,\bar R+1)}{(\bar R-1)^4}+
\frac{R\,(R^2+6\,R+1)}{(R-1)^4}\;\frac{(\bar R+1)^2}{(\bar R-1)^2}\bigg)
\nonumber\\
&&+256\,\frac{R^{\frac32}(R+1)}{(R-1)^4}\;\frac{\bar R^{\frac32}\,(\bar R+1)}{(\bar R-1)^4}+
16\,\frac{R\,(R^2+6\,R+1)}{(R-1)^4}\;\frac{\bar R\,(\bar R^2+6\,\bar R+1)}{(\bar R-1)^4}
\Bigg\}.
\end{eqnarray}
Note that this is the correct form for the 4-point function for weights (2,2), (1,1), (1,1), (2,2).

\subsection{Coincidence limit and operator mixing}\label{stlimit}
We investigate the operator mixing by taking the coincidence limit $(a_1,\bar a_1)\to(0,0)$ and
$(a_2,\bar a_2)\to (b,\bar b)$.
Similar to the case of the dilaton in section (\ref{opmxdil}), there are singularities which
correspond to mixing with quasi-primary operators with half-integer conformal weights.
All the descendants of these fields also have half-integer weights.
This type of mixing contributes to the wave function renormalization (\ref{renorm}).
It does not  affect the anomalous dimension of the string state which has conformal
weight $(2,2)$.
Under the above coincidence limit, the singular part of the four-point function which
corresponds to mixing with integer weight operators is of the form
\begin{eqnarray}\label{stcl}
&&2^{-12}\Bigg\{
\frac{81}{4}\frac{1}{a_1^2\,(a_2-b)^2\,\bar a_1^2\,(\bar a_2-\bar b)^2\,|b|^4}+\,
\frac{81}{2}\,\frac{\Big(a_1\,\bar b-(a_2-b)\,\bar b+\bar a_1\,b-(\bar a_2-\bar b)\,b\Big)}
{a_1^2\,(a_2-b)^2\,\bar a_1^2\,(\bar a_2-\bar b)^2\,|b|^6}\nonumber\\
&&+\,\frac{27}{a_1^2\,(a_2-b)^2\,\bar a_1^2\,(\bar a_2-\bar b)^2\,|b|^8}\,
\Big(-5\,a_1\,(a_2-b)\,\bar b^2-5\,\bar a_1\,(\bar a_2-\bar b)\,b^2\nonumber\\
&&+\,3\,a_1\,\bar a_1\,|b|^2-3\,a_1\,(\bar a_2-\bar b)\,|b|^2
-3\,(a_2-b)\,\bar a_1\,|b|^2+3\,(a_2-b)\,(\bar a_2-\bar b)\,|b|^2\Big)\nonumber\\
&&+\,\frac{54}{a_1^2\,(a_2-b)^2\,\bar a_1^2\,(\bar a_2-\bar b)^2\,|b|^{10}}\,
\Big(5\,a_1\,(a_2-b)\,(\bar a_2-\bar b)\,b\,\bar b^2
-5\,a_1\,(a_2-b)\,\bar a_1\,b\,\bar b^2\nonumber\\
&&-\,5\,a_1\,\bar a_1\,(\bar a_2-\bar b)\,b^2\,\bar b
+5\,(a_2-b)\,\bar a_1\,(\bar a_2-\bar b)\,b^2\,\bar b\Big)\nonumber\\
&&+\,\frac{900}{a_1\,(a_2-b)\,\bar a_1\,(\bar a_2-\bar b)\,|b|^8}+\cdots\Bigg\},
\end{eqnarray}
where $``\cdots"$ corresponds to subleading singularities.
We can now move on to subtracting the relevant conformal family.

\subsection{Conformal family subtraction and mixing coefficients}\label{stsubtract}
Let us consider the leading singular term in the above expansion with
the coefficient 81/4.
According to (\ref{ope}), this term corresponds to operator mixing with a quasi-primary,
or a linear combination of quasi-primaries, with conformal weight $(1,1)$.
Charge conservation requires this operator to be a singlet under
the R-symmetry $SU(2)_L\times SU(2)_R$ and the internal symmetry
$SU(2)_1\times SU(2)_2$.
We show that the weight (1,1) operator is indeed the deformation operator (\ref{defool}).
To see this, we evaluate the three-point function
\begin{equation}\label{3pfmx1}
\lim_{y,\bar y\to\infty}
y^{2h_{d}}\bar y^{2\tilde h_d}
\Big<\sum_{\kappa=1}^{2}\phi_{\mathrm{st}\,(\kappa)}(a_1,\bar a_1)\;\;
\lambda\mathcal{O}_d
(0,0)\;\;
\lambda\mathcal{O}_d
(y,\bar y)
\Big>=\frac{C_{iAA}}
{a_1^{h_{\mathrm{st}}+h_d-h_d}\,\bar a_1^{\tilde h_{\mathrm{st}}+\tilde h_d-\tilde h_d}}
\end{equation}
where the indices $i$ and $A$ in $C_{iAA}$ correspond to $\phi_{\mathrm{st}}$ and the
deformation, respectively.
$(h_{\mathrm{st}},\tilde h_{\mathrm{st}})$ and $(h_d,\tilde h_d)$
are the conformal weights of $\phi_{\mathrm{st}}$ and the deformation, respectively.
The above three-point functioned, normalized by two-point functions, is
\begin{equation}\label{3pfCiAA}
\frac{C_{iAA}}
{a_1^{h_{\mathrm{st}}+h_d-h_d}\,\bar a_1^{\tilde h_{\mathrm{st}}+\tilde h_d-\tilde h_d}}=
\frac{18}{2^8}\,\frac{1}{|a_1|^4},
\end{equation}
which then gives $C_{iAA}=9\times2^{-7}$.
Since we have normalized the operators by their two-point functions, the coefficient
of the three-point function and the coefficient of the most singular term in the OPE
(\ref{ope})
are equal: $C_{iAA}=C^{A\,\{0,0\}}_{iA}$.
Denoting the complete deformation operator as  $O_A$, we find the operator algebra
\begin{equation}\label{ope2}
\sum_{\kappa=1}^{2}\phi_{\mathrm{st}\,(\kappa)}(a_1,\bar a_1)\,
O_A(0,0)=
\sum_{\{k,\tilde k\}}
C^{A\,\{k,\tilde k\}}_{iA}a_1^{-2+K}\,\bar a_1^{-2+\tilde K}
O_A^{\{k,\tilde k\}}(0,0),
\end{equation}
where the deformation operator is the ancestor of this conformal family.
The operator algebra for the other two operators
$\sum_{\kappa=1}^{2}\phi_{\mathrm{st}\,(\kappa)}(a_2,\bar a_2)$ and
$O_A(b,\bar b)$ is obtained in a similar way.
Inserting the two OPEs in the four-point function (\ref{stAAsttot}) we obtain the two-point
function
\begin{eqnarray}\label{4pf2pf}
\Bigg<
\bigg(\sum_{\{k^\prime,\tilde k^\prime\}}
C^{\prime\,A\,\{k^\prime,\tilde k^\prime\}}_{iA}
(a_2-b)^{-2+K^\prime}\,(\bar a_2-\bar b)^{-2+\tilde K^\prime}
O_A^{\{k^\prime,\tilde k^\prime\}}\bigg)(b,\bar b)\times\;
\qquad\qquad\qquad\quad
\nonumber\\
\bigg(
\sum_{\{k,\tilde k\}}C^{A\,\{k,\tilde k\}}_{iA}a_1^{-2+K}\,\bar a_1^{-2+\tilde K}
O_A^{\{k,\tilde k\}}
\bigg)(0,0)
\Bigg>.
\end{eqnarray}
Let us first consider the most singular terms of the two OPEs which corresponds
to the ancestor field with $\{k,\tilde k\}=\{k^\prime,\tilde k^\prime\}=\{0,0\}$.
The above two-point function then reads
\begin{eqnarray}\label{4pf2pf-00}
&&\Bigg<
\bigg(\frac{C^{\prime\,A\,\{0,0\}}_{iA}}{(a_2-b)^{2}\,(\bar a_2-\bar b)^{2}}\,
O_A^{\{0,0\}}\bigg)(b,\bar b)\;\;
\bigg(\frac{C^{A\,\{0,0\}}_{iA}}{a_1^{2}\;\bar a_1^{2}}\,
O_A^{\{0,0\}}\bigg)(0,0)
\Bigg>\nonumber\\
&&=\frac{1}{2^{12}}\,\frac{81}{4}\,\frac{1}{|a_1|^4\,|a_2-b|^4\,|b|^4}.
\end{eqnarray}
This gives the leading singular term that we found in the expansion (\ref{stcl}) with
exactly the same coefficient.
Therefore, the deformation operator accounts for the leading singularity of the
four-point function.

We will next have to compute the contribution of the descendant operators in
(\ref{4pf2pf}) and subtract them from the subleading singular terms in the expansion
(\ref{stcl}).
We will only need to find the contribution of the descendants $O_A^{\{1,0\}}$,
$O_A^{\{0,1\}}$, and $O_A^{\{1,1\}}$, since higher descendants have conformal
weights larger that $(2,2)$ and play no role in determining the anomalous dimension
of the string state.
Computation of the structure constants of the descendant fields is explained in
\cite{DiFrancesco:1997nk}.
In general, in the operator algebra of two quasi-primary fields $O_1$ and $O_2$
with conformal weights $h_1$ and $h_2$,
\begin{equation}
O_1(z,\bar z)\,O_2(0,0)=\sum_p\sum_{\{k,\tilde k\}}
C^{p\,\{k,\tilde k\}}_{12}z^{h_p-h_1-h_2+K}\,
\bar z^{\tilde h_p-\tilde h_1-\tilde h_2+\tilde K}O_p^{\{k,\tilde k\}}(0,0),\nonumber
\end{equation}
the structure constants of the descendants $C^{p\,\{k,\tilde k\}}_{12}$ are determined
by the structure constant of the ancestor field through the relation
\begin{equation}\label{struc}
C^{p\,\{k,\tilde k\}}_{12}=C^{p\,\{0,0\}}_{12}\,
\beta_{12}^{p\,\{k\}}\,\tilde\beta_{12}^{p\,\{\tilde k\}},
\end{equation}
where $\beta_{12}^{\{k\}}$ and $\tilde\beta_{12}^{\{\tilde k\}}$ are coefficients which
depend only on the central charge and the conformal weights.
For the descendant states at level 1 the coefficients are given by:
\begin{equation}\label{beta}
\beta_{12}^{p\{1\}}=\frac{h_p+h_1-h_2}{2\,h_p},\qquad
\tilde\beta_{12}^{p\,\{1\}}=\frac{\tilde h_p+\tilde h_1-\tilde h_2}{2\,\tilde h_p}.
\end{equation}
In our case, $h_p=1$, $h_1=2$, and $h_2=1$ and we have:
\begin{equation}\label{betast}
\beta_{iA}^{A\{1\}}=1,\qquad\tilde\beta_{iA}^{A\,\{1\}}=1.
\end{equation}
The structure constants then read
\begin{equation}\label{strucst}
C^{A\,\{1,1\}}_{iA}=C^{A\,\{1,0\}}_{iA}=C^{A\,\{0,1\}}_{iA}=C^{A\,\{0,0\}}_{iA}=\frac{9}{2^7}.
\end{equation}
We can now calculate the contribution of the descendant fields $O_A^{\{k,\tilde k\}}(0,0)$
and $O_A^{\{k^\prime,\tilde k^\prime\}}(b,\bar b)$ in (\ref{4pf2pf}).
As mentioned earlier, for each OPE, we are interested in the three level 1 descendants
($O_A^{\{1,0\}},$ $O_A^{\{0,1\}},$ and $O_A^{\{1,1\}}$).
Therefore there are nine terms to evaluate.
For example, for $\{k,\tilde k\}=\{1,0\}$ and $\{k^\prime,\tilde k^\prime\}=\{0,0\}$, we have
$K=1$, $\tilde K=K^\prime=\tilde K^\prime=0$, and obtain
\begin{eqnarray}\label{4pf2pf-10}
&&\Bigg<
\bigg(\frac{C^{\prime\,A\,\{0,0\}}_{iA}}{(a_2-b)^{2}\,(\bar a_1-\bar b)^{2}}\,
O_A^{\{0,0\}}\bigg)(b,\bar b)\;
\bigg(\frac{C^{A\,\{1,0\}}_{iA}}{a_1\;\bar a_1^{2}}\,
O_A^{\{1,0\}}\bigg)(0,0)
\Bigg>=\nonumber\\
&&\frac{1}{2^{12}}\,\frac{81}{4}\,
\frac{1}{a_1\,(a_2-b)^2}\,\frac{1}{\bar a_1^2\,(\bar a_2-\bar b)^2}\,L_{-1}\,
\langle O_A^{\{0,0\}}(b,\bar b)\;O_A^{\{0,0\}}(0,0)\rangle=\nonumber\\
&&\frac{1}{2^{12}}\,\frac{81}{4}\,
\frac{1}{a_1\,(a_2-b)^2}\,\frac{1}{\bar a_1^2\,(\bar a_2-\bar b)^2}\,\partial_z
\langle O_A^{\{0,0\}}(b,\bar b)\;O_A^{\{0,0\}}(z,\bar z)\rangle\Big|_{z\to0}=\nonumber\\
&&\frac{1}{2^{12}}\,\frac{81}{2}\,
\frac{a_1\,\bar b}{a_1^2\,(a_2-b)^2\,\bar a_1^2\,(\bar a_2-\bar b)^2\,|b|^6}.
\end{eqnarray}
We perform similar computations for the eight remaining terms and subtract them
from the corresponding singularities in the expansion (\ref{stcl}).
The remaining singular terms are
\begin{eqnarray}\label{stcl-sub1}
&&2^{-12}\Bigg\{
\,\frac{27}{a_1^2\,(a_2-b)^2\,\bar a_1^2\,(\bar a_2-\bar b)^2\,|b|^8}\,
\Big(-\frac{1}{2}\,a_1\,(a_2-b)\,\bar b^2
-\frac{1}{2}\,\bar a_1\,(\bar a_2-\bar b)\,b^2\Big)\nonumber\\
&&+\,\frac{27}{a_1^2\,(a_2-b)^2\,\bar a_1^2\,(\bar a_2-\bar b)^2\,|b|^{10}}\,
\Big(a_1\,(a_2-b)\,(\bar a_2-\bar b)\,b\,\bar b^2
-\,a_1\,(a_2-b)\,\bar a_1\,b\,\bar b^2\nonumber\\
&&-\,a_1\,\bar a_1\,(\bar a_2-\bar b)\,b^2\,\bar b
+\,(a_2-b)\,\bar a_1\,(\bar a_2-\bar b)\,b^2\,\bar b\Big)\nonumber\\
&&+\,\frac{171}{a_1\,(a_2-b)\,\bar a_1\,(\bar a_2-\bar b)\,|b|^8}+\cdots\Bigg\}.
\end{eqnarray}
The first line of the above equation has two terms in it:
\begin{eqnarray}
-\frac{27}{2}\,\frac{1}{a_1\,(a_2-b)\,\bar a_1^2\,(\bar a_2-\bar b)^2\,b^4\,\bar b^2},
\label{st21mixing}\\
-\frac{27}{2}\,\frac{1}{a_1^2\,(a_2-b)^2\,\bar a_1\,(\bar a_2-\bar b)\,b^2\,\bar b^4}.
\label{st12mixing}
\end{eqnarray}
The first term (\ref{st21mixing}) shows that there is operator mixing with a quasi-primary
or a linear combination of quasi-primaries with conformal weight $(2,1)$.
These fields are the ancestors of their conformal families.
The second term (\ref{st12mixing}) signals mixing with quasi-primaries with conformal
weight $(1,2)$.
We can use the conformal algebra of the theory and construct quasi-primary operators
with the required conformal weights which contribute to these mixings.
In our case, however, we do not need to identify explicitly all the $(2,1)$ or $(1,2)$
operators which mix with $\phi_{\mathrm{st}}$.
We can indeed evaluate the structure constants of all the descendant fields
using (\ref{struc}).
Thus the only information needed are the coefficients $\beta_{12}^{p\,\{k\}}$ and
$\tilde\beta_{12}^{p\,\{\tilde k\}}$.
We are only interested in the level one descendants which have conformal weights
$(2,2)$.
The values of $\beta_{12}^{p\,\{1\}}$ and $\tilde\beta_{12}^{p\,\{\tilde 1\}}$ are therefore
the same as above (\ref{betast}): $\beta_{12}^{p\,\{1\}}=\tilde\beta_{12}^{p\,\{\tilde 1\}}=1$.
Following similar computations as in the previous case we compute the contributions of
the descendant fields and subtract them from the corresponding singular terms in
(\ref{stcl-sub1}).
The remaining singularity is of the form:
\begin{eqnarray}\label{stcl-sub2}
2^{-12}\,\frac{9}{a_1\,(a_2-b)\,\bar a_1\,(\bar a_2-\bar b)\,|b|^8}+\cdots.
\end{eqnarray}
This shows that there is a quasi-primary or a linear combination of quasi-primary
operators which have conformal dimension $(2,2)$ and mix with our candidate
operator $\phi_{\mathrm{st}}$ at the first order in perturbation theory.
Since these quasi-primaries have the same weight as $\phi_{\mathrm{st}}$,
they contribute to the anomalous dimension of the candidate operator at the first order.
Operators with different weights which mix with $\phi_{\mathrm{st}}$ will
contribute to the wave function renormalization.

We can see the above results in an alternative way,
using a trick that is available for this case.
In the coincidence limit we considered above, we set $a_1,\bar a_1\to0$ and
$a_2,\bar a_2\to b,\bar b$.
We can equally well send the deformation operator $O_A$ to the vicinity of
$\phi_\mathrm{st}$.
Let us assume that the two deformations are at positions $z=z_1$ and $z=z_2$
in the base space.
We take the coincidence limit $(z_1,\bar z_1)\to(a_1,\bar a_1)$,
$(z_2,\bar z_2)\to(a_2,\bar a_2)$ and then set $z_1=0$, $z_2=b$.
Under this coincidence limit, the singular part of the four-point correlation function
(\ref{stAAsttot}) is of the form
\begin{eqnarray}\label{stcl-alt}
2^{-12}\Bigg\{
\frac{81}{4}
\frac{1}{a_1^2\,(a_2-b)^2\,\bar a_1^2\,(\bar a_2-\bar b)^2\,|a_1-a_2|^4}
\qquad\qquad\qquad\qquad
\nonumber\\
-\,\frac{27}{2}
\frac{\Big(a_1\,(a_2-b)\,(\bar a_1-\bar a_2)^2+\bar a_1\,(\bar a_2-\bar b)\,(a_1-a_2)^2\Big)}
{a_1^2\,(a_2-b)^2\,\bar a_1^2\,(\bar a_2-\bar b)^2\,|a_1-a_2|^8}
\quad
\nonumber\\
+\,\frac{9}{a_1\,(a_2-b)\,\bar a_1\,(\bar a_2-\bar b)\,|a_1-a_2|^8}+\cdots\Bigg\}.
\qquad\qquad\quad\;
\end{eqnarray}
The first line shows operator mixing with a $(1,1)$ quasi-primary.
As was shown in (\ref{4pf2pf-00}), this operator is the deformation operator, which
accounts for the leading singularity.
We note that in the present coincidence limit we have $h_1=1$ and $h_2=2$ in the
operator algebra (\ref{ope}).
Therefore, the coefficients $\beta_{12}^{p\,\{1\}}$ and $\tilde\beta_{12}^{p\,\{\tilde 1\}}$
(\ref{beta}) for the level 1 descendants of the deformation operator are now
\begin{equation}\label{betast-alt}
\beta_{iA}^{A\{1\}}=0,\qquad\tilde\beta_{iA}^{A\,\{1\}}=0.
\end{equation}
Hence, the structure constants of these descendants vanish and only the ancestor
of the family contributes to the expansion.
The second line of equation (\ref{stcl-alt}) then tells us that there are quasi-primaries
with weight $(2,1)$ and $(1,2)$ which contribute to operator mixing.
The coefficient of this singularity agrees with that obtained in (\ref{stcl-sub1}).
Again, the structure constants of the descendants of these operators vanish and there
will be no contribution from these descendants to the expansion.
Finally, the last line in (\ref{stcl-alt}) implies that there is at least one $(2,2)$ quasi-primary
operator which mixes with $\phi_{\mathrm{st}}$ at the first order.
The coefficient agrees with (\ref{stcl-sub2}).
While this shortcut is available for this computation, this is not always the case. The earlier procedure is generically applicable.

In order to evaluate the anomalous dimension to the first order we need to determine
all $(2,2)$ quasi-primary operators that contribute to the operator mixing in (\ref{stcl-sub2}).
We then need to determine all the $(2,2)$ quasi-primaries that mix with these operators.
We have to continue this search until we find all such $(2,2)$ operators.
We will then be able to diagonalize the matrix of the structure constants and find the
anomalous dimension of our candidate string state.

\section{Summary and outlook}
\label{discsect}

In this paper and the companion work \cite{companion}, we began an investigation into how the anomalous dimensions of low-lying string states in the D1-D5 SCFT are lifted as we perturb away from the orbifold point where string calculations are easiest to do. We found evidence of operator mixing at first order, which means that as we increase the deformation parameter the anomalous dimensions of some of the string states will head downwards while others will head upwards.

Our method starts with evaluating four-point functions involving the operator of interest, the deformation operator, and their Hermitean conjugates. The two deformation operator insertions are of course required because we want to perturb away from the orbifold point (towards the gravity limit). Using four-point functions may seem a tad roundabout, but it is actually more efficient than starting with three-point functions. The reason is that we can use factorization channels of four-point functions to identify which intermediate quasiprimary operators participate in mixing. This cuts down on the number of independent three-point functions we need to calculate. Once we know the conformal weight of such an intermediate quasiprimary, we can subtract its conformal family from the leading order singular limit of the four-point function. Such a family contains an ancestor quasi-primary operator and all its descendents under the Virasoro algebra. After subtraction, some leading order singularities in coincidence limits will typically remain. We will therefore continue iterating the procedure until we exhaust all the intermediate conformal families which mix with our original operator. Diagonalizing the resulting matrix of structure constants will eventually yield the anomalous dimensions that we wanted in the first place, at leading order in perturbation theory.

In general, solving for string state anomalous dimensions is an extremely hard problem. We need the iteration procedure outlined above to truncate, in order to be able to find the anomalous dimensions for the string states we are after. Diagonalizing an infinite-dimensional matrix of structure constants, after all, would be practically impossible.
The key observation is that for low-lying string states the number of
intermediate quasi-primaries which can mix with them should be finite.
This is why we have focused on a low-lying string state which is about as simple as possible (but no simpler): $\partial X \partial X \bar{\partial} X {\bar{\partial}}X$.

In order to be able to compute the four-point and three-point functions we needed, we had to further develop Lunin-Mathur symmetric orbifold technology. In particular, we needed to know how lifting to the covering space worked, including all the details of twist and nontwist operators, fractional modings, ramification points, images, bosonization of fermions, and so forth. We illustrated our developments of Lunin-Mathur symmetric orbifold technology in \cite{companion} with a simple example and with a more complicated one involving excitations, fermions, and currents. In section \ref{coc} of this work, we also found a suitable representation of cocycles transforming correctly under $SU(2)_L\times SU(2)_R$ R-symmetry and the internal $SU(2)_1\times SU(2)_2$, which involved features that we had not seen elsewhere.

We have several items on our remaining to-do list. Of highest priority is to enumerate {\em all} the operators that mix with $\partial X \partial X {\bar{\partial}} X {\bar{\partial}}X$ -- and the operators
that they mix with in turn. Once we have that complete list of operators, we can subtract their conformal families, iterating our procedure until all of the mixing coefficients are nailed down. This will permit us to diagonalize the matrix of structure constants and find the desired anomalous dimensions. We are also investigating {other} choices of low-lying string states coming from both the non-twist and the twist sector.
Our other immediate goal is to connect with the work \cite{Gava:2002xb} of Gava and Narain, who also investigated anomalous dimensions of particular low-lying string states, in the context of proving pp-wave/CFT$_2$ duality. The string states they considered are right-chiral; the left sector has excitations with
fractional modes of conserved currents. They analyzed the anomalous dimension for a class of states by calculating three-point functions of the CFT. They found that it is proportional to $(k/n)^2$, where $n$ is the twist order and $k$ gives the fractional mode number. In order to compute the anomalous dimension of more general string states with excitations on both the left and right sides, it is necessary to compute full four-point functions of the CFT. We plan to investigate this directly by using the methods of this paper.

\section*{Acknowledgements}

The authors wish to thank Samir Mathur for ideas, interesting discussions, and guidance, and for hospitality during IGZ's visits.
We also wish to thank Steven Avery for helpful discussions.

AWP wishes to thank KITP at UCSB for support during the ``Bits `n' Branes" workshop. IGZ is grateful to the Simons Center for Geometry and Physics for support during the {``Superconformal Theories in Diverse Dimensions"} workshop, and to Shlomo Razamat and Cumrun Vafa for helpful discussions.

This research was supported by the Canadian Institute of Particle Physics (IPP) and the Natural Sciences and Engineering Research Council (NSERC) of Canada.

\pagebreak

\end{document}